%% file: klessen.tex
\documentclass[usenatbib]{article}
\usepackage{natbib}
\usepackage{graphicx}
\usepackage{amsmath}
\usepackage{amssymb}
\usepackage[pdfpagelabels=false]{hyperref}  

\begin{document}

\input{klessen-ch4}

\bibliographystyle{ws-book-har}    
\bibliography{klessen-ref}      

\end{document}

%% file: klessen-ch4.tex
{
\Huge \bf Formation of the first stars \label{chap:FirstStars}
}

\vspace*{1.1cm}
{
\Large\sf Ralf S.\ Klessen\\[0.15cm]
Universit\"{a}t Heidelberg, Zentrum f\"{u}r Astronomie\\[0.15cm]
Albert-Ueberle-Str. 2, 69120 Heidelberg, Germany
}
\vspace*{0.5cm}

{\large
To be published in 
the review volume {\em Formation of the First Black Holes}, \\[0.1cm]
editors M.\ Latif and D.\ R.\ G.\ Schleicher, pages 67 -- 98 \\[0.1cm] 
World Scientific Publishing Company, 2018 }

\vspace*{1.9cm}

{\large \bf Abstract}\\
From studying the cosmic microwave background, we know our Universe started out very simple. It was by and large homogeneous and isotropic, with small fluctuations that can be described by linear perturbation theory. In stark contrast, the Universe today is highly structured on a vast range of length and mass scales. In the evolution towards increasing complexity, the formation of the first stars marks a primary transition event. The first generation of stars, the so-called Population III (or Pop. III) build up from truly metal-free primordial gas. They have long been thought to live short, solitary lives, with only one massive star forming per halo. However, in recent years this simple picture has undergone substantial revision, and we now understand that stellar birth in the early Universe is subject to the same complexity as star formation at present days. In this chapter, I review the current state of the field. I begin by introducing the basics concepts of star-formation theory and by discussing the typical environment in which Pop. III stars are thought to form. Then I argue that the accretion disk that builds up in the center of a halo is likely to fragment, resulting in the formation of a cluster of stars with a wide range of masses, and I speculate about how this process may be influenced by stellar feedback, the presence of magnetic fields, the energy input from dark matter annihilation, and the occurrence of large-scale streaming velocities between baryons and dark matter. Finally, I discuss direct and indirect constraints on Pop. III star formation from high-redshift observations and from the search for extremely metal-poor stars in the Milky Way and its satellites. 

\newpage
\section{Introduction}
\label{sec:FS-Intro}

The appearance of the first stars marked a primary transition in cosmic history. Their light ended the so-called `dark ages', and they played a key role in cosmic metal enrichment and reionization, thereby shaping the galaxies and their internal properties as we see them today. Understanding high-redshift star formation is thus important for many areas of modern astrophysics. However, studying stellar birth in the primordial Universe is a relatively young area of astrophysical sciences. Only with the advent of new numerical methods and powerful supercomputers did the numerical modeling of early star formation become feasible. As a consequence, there is still little consensus on the physical processes that govern stellar birth at high redshifts. This chapter aims at providing a concise overview of the current state of the field.

The first generation of stars, the so-called Population III (or Pop. III) built up from truly metal-free primordial gas. They have long been thought to live short, solitary lives, with only one extremely massive star forming in each dark matter halo with about 100 solar masses or more  \citep{Omukai2001,Abel2002,Bromm2002,Oshea2007}. The idea was that first star formation is simple and one only needs to know the initial Gaussian density perturbations of material at very high-redshift which are very well understood, e.g.\ from measuring the cosmic microwave background, see  \citet{Planck2016}, the growth of cosmological structures, and  the heating and cooling processes in the primordial gas, which are thought to be very simple because of the lack of elements heavier than hydrogen and helium. In addition, the early numerical models had considerable limitations. First, they only followed the collapse of the very first object in the first halo to form, since the hydrodynamic timestep decreases as the gas collapses, and so the simulations effectively grind to a halt once the protostar has formed. Second, most calculations did not include the effects of protostellar feedback, which may substantially reduce the accretion rate onto the protostar by dumping energy and/or momentum into the infaling gas. 

The ever increasing capabilities of modern supercomputers allow us now to perform significantly improved numerical simulations of early star formation that reveal a much more complex picture. The introduction of numerical techniques to first star formation studies that have been standard repertoire in present-day star formation lead to the conclusion that fragmentation is a wide-spread phenomenon in first star formation \citep{Clark2011b, Greif12}, and that Pop. III stars form as members of multiple stellar systems with separations as small as the distance between the Earth and the Sun, see \citet{Turk09, Clark11, Greif2011, Smith11, Stacy2013}. Studies that do include radiative feedback \citep{Hirano2014, Hirano2015, Hosokawa2016}, magnetic fields \citep{Machida2006, Machida2008, Schleicher2009, Schliecher2010dyn, Sur2010, Sur2012, Turk2012, Schobera, Schoberb, Bovino2013NJP}, dark matter annihilation \citep{Smith2012, Stacy2014}, as well as the primordial streaming velocities \citep{Tseliakhovich2010, Greif2011, Maio11, Stacy2011} add to this complexity. All of these processes are relevant and need to be included in any realistic model. There is agreement now that primordial star formation is equally dynamic and difficult to understand as stellar birth at present days.

For further reading on primordial star formation, we refer to the reviews of  \citet{Bromm04,Yoshida2012,Bromm2013}, or to the book by \citet{Loeb2010}. A good overview with specific focus on the astrochemistry of first star formation is provided by \citet{Glover2005, Glover2013}. The current numerical frontier of high-redshift star formation is summarized by \citet{Greif15}. For further reading on the transition to the second generations of stars and build up of the first galaxies we recommend \citet{Bromm2011A} or the textbook by \citet{Loeb2013}. As we calibrate our understanding of primordial star formation with what we know about stellar birth at present days, we also recommend the reviews by \citet{Maclow2004, Mckee2007,Zinnecker2007, Padoan2014,Krumholz2015, Klessen2016} on different aspects of this subject.

In this chapter we try to provide a brief summary of primordial star formation theory, we discuss some of the difficulties we face in this context and speculate about possible observational constraints from the high-redshift Universe as well as from our local neighborhood in the Galaxy. Next, in Section~\ref{sec:basics} we introduce some of the basic concepts that are behind any theory of star formation. The environment of Pop. III star formation and the thermodynamic behavior of primordial gas are discussed in Section~\ref{sec:environment}. Section~\ref{sec:complexity} illustrates the complexity of stellar birth in the early Universe. We start with simple one-dimensional collapse models, but then argue that the accretion disks around the first stars are highly susceptible to fragmentation. We also speculate about how stellar feedback, magnetic fields, possible dark matter annihilation,  and  large-scale streaming velocities between baryons and dark matter may influence the stellar mass spectrum. In Section~\ref{sec:observations} we discuss indirect constraints on the properties of the first stars from high-redshift observations and from stellar archeology in the solar neighborhood, and we speculate about the possible detection of genuine low-mass Pop. III stars in the Universe today. We  summarize and conclude  in  Section~\ref{sec:summary}.

\section{Basic concepts}
\label{sec:basics}
The mean density of the Sun is $1.4\,$g$\,$cm$^{-3}$, and the numbers for other stars are very similar \citep{Kippenhahn2012}. In contrast to that, the mean density of the gas in the Milky Way at the solar radius is $\sim 10^{-23}\,$g$\,$cm$^{-3}$ \citep{Feri2007}, and the mean baryon density of the Universe at present days is with $5 \times 10^{-31}\,$g$\,$cm$^{-3}$ even less \citep{Planck2016}. The formation of stars therefore requires a density increase by many orders of magnitude. The only way nature can achieve such enormous density contrasts is by gravitational collapse. All other fundamental forces are either short range, as the strong and weak nuclear forces, or cancel out on average. The electromagnetic force plays no role on global scales because the Universe as a whole is charge neutral and so there is no net force between positive and negative charges. And so it is by far the weakest of the four fundamental forces, gravity, that dominates the large-scale dynamics, simply because it is the only one that is purely attractive and has infinite reach.  

\subsection{Stability of spherical gas clouds -- Jeans criterion}
\label{subsub:Jeans}

As argued above, any theory of star formation needs to be based on the competition between gravitational attraction and a large number of dispersing processes ranging from gas pressure, to the coupling between matter and radiation, magnetic fields, turbulence, cosmic rays to name some of the most important ones \citep{Maclow2004, Mckee2007}. In its most basic form, a criterion for gravitational collapse goes back to \citet{Jeans1902}, who studied the stability of self-gravitating isothermal gas spheres. He found that these systems have a critical mass,
\begin{eqnarray}
M_{\rm J} &=& \dfrac{\pi^{5/2}}{6}\left( \dfrac{1}{G}\right)^{3/2} \rho^{-1/2} c_{\rm s}^{3} \label{eq:Jeans-mass}
\\ 
&=&\dfrac{\pi^{5/2}}{6} \left( \dfrac{k}{G}\right)^{3/2} \left(\dfrac{1}{\mu m_{\rm H}}\right)^{2} n^{-1/2} T^{3/2} \nonumber\\ 
&\approx& 50 \,{\rm M}_{\odot} \;\mu^{-2} \left(\frac{n}{1 {\rm cm}^{-3}}\right)^{-1/2} \left(\frac{T}{1{\rm K}}\right)^{3/2}\;, \nonumber
\end{eqnarray}
where the proportionality factor depends on Boltzmann's constant $k$, on the gravitational constant $G$, on the proton mass $m_{\rm H}$, and on the mean molecular weight of the gas particles $\mu$. Recall that $\mu \approx 1.22$ for primordial atomic gas, and that $\mu \approx 2.33$ for molecular gas.  If this critical mass is exceeded, gravity overwhelms pressure gradients and the systems collapses. If the mass is smaller, then gas pressure wins and the sphere expands. The Jeans mass $M_{\rm J}$ depends inversely on the square root of the density $\rho$ and on the third power of the sound speed $c_{\rm s}$. Because of $\rho = \mu m_{\rm H} n$ and $c_{\rm s} = (kT/\mu m_{\rm H})^{1/2}$, we can  also write this expression in term of the number density $n$ and the temperature $T$ as $M_{\rm J} \propto n^{-1/2} T^{3/2}$. In summary, the larger the density the smaller is the critical mass for collapse, and the higher the temperature the more stable is the system. 

The concept of the Jeans mass (\ref{eq:Jeans-mass}) can be extended to include additional physics by introducing an 'effective' density (say when dealing with multi-component fluids) and an 'effective' temperature or sound speed. For example when the gas is turbulent on scales much smaller than the dynamical scales of interest \citep{Chandrasekhar1951A, Chandrasekhar1951B, VonWeiz1951}  then the velocity dispersion $\sigma$ can be simply added to the sound speed,  
\begin{equation}
c_{\rm s, eff} = (c_{\rm s}^2 + \sigma^2)^{1/2}\;.
\label{eq:effective-sound-speed}
\end{equation} 
Similar has been proposed for the presence of magnetic fields, then we add half of the Alfv\'{e}n velocity squared, $v_{\rm A}^2 = B^2/4\pi\rho$, in Equation (\ref{eq:effective-sound-speed}), see e.g.\ \citet{Federrath2012}. In order to trigger star formation in an otherwise stable medium, either the density needs to increase (say, due to an external compression) or the temperature needs to decrease (by some cooling process). Consequently, a significant fraction this chapter focuses on a discussion of the various astrophysical processes that can lead to this behavior. 

We note that the Jeans criterion can also be formulated in terms of energies. If the absolute value of the potential energy $W$ is more than twice the integral over the pressure $U$, which is closely related to the internal energy of the system, then collapse occurs. We get
\begin{equation}
\eta_{\rm vir} = \dfrac{|W|}{U} = \dfrac{\frac{1}{2}\int d^3x \rho(x) \phi(x)}{\frac{3}{2}\int d^3x P(x)} > 1/2\;,
\end{equation}
where $\phi$ is the potential and $P$ is the pressure. This is simply the scalar virial theorem. Another alternative is to look at typical timescales. If the free-fall time, $\tau_{\rm ff}$, is smaller than the sound crossing time, $\tau_{\rm s}$, then the system collapses. Otherwise sound waves travel fast enough to wipe out inhomogeneities and the system is stable. Both times are defined as 
\begin{equation}
\tau_{\rm ff} = \left( \dfrac{3\pi}{32 G \rho} \right)^{1/2} ~~~{\rm and}~~~~\tau_{\rm sound} =  \dfrac{R}{c_{\rm s}}\;,
\label{eq:timescales}
\end{equation}
where $R$ is the radius of the sphere. 

We can obtain a crude estimate of the accretion rate onto the center of the halo from combining Equations (\ref{eq:Jeans-mass}) and (\ref{eq:timescales}) as 
\begin{equation}
\dot{M} =\, \zeta \dfrac{M_{\rm J}}{\tau_{\rm ff}} \propto \dfrac{c_{\rm s}^{3}}{G}\;,
\label{eq:accretion-rate}
\end{equation}
where the factor $\zeta$ can take values of up to several tens depending on the actual density profile and on how much the gas mass $M$ exceeds the Jeans mass $M_{\rm J}$, for the classical studies, see \citet{Shu1977,Larson1969, Penston1969, Whitworth1985}. Note that the accretion rate for isothermal collapse only depends on the gas temperature.

\subsection{Stability of rotating disks -- Toomre criterion}
\label{subsec:Toomre}

In systems that are supported by rotational motions, like protostellar accretion disks or spiral galaxies,  the criterion for gravitational instability takes a slightly different form. Besides gas pressure now also the stabilizing effect of shear in the disk needs to be taken into account \citep{Toomre1964}. The criterion for instability reads
\begin{equation}
Q = \dfrac{\kappa c_{\rm s}}{\pi G \Sigma} \lesssim 1\;,
\label{eq:Toomre}
\end{equation}
with the surface density $\Sigma$ and the epicyclic frequency $\kappa$, which for Keplerian disks takes the simple form $\kappa = \Omega$, where $\Omega$ is the rotational frequency, see  the review by \citet{Kratter2016}. The criterion was originally derived for infinitesimally thin disks, but it can be extended to thick disks with multiple components \citep{Rafikov2001, Elmegreen2002} which introduces correction factors of order unity to Equation (\ref{eq:Toomre}). Again, there are two main pathways towards disk fragmentation. If the mass load onto the disk by accretion from the enclosing gas cloud exceeds its capability to transport material inwards by viscous stresses, $\Sigma$ increases beyond the critical value and the disk becomes unstable. First, spiral arms form and speed up the inward transport. If this is not enough, these arms become non-linear and interact with each other with run-away collapse occuring in the interception regions.
By a similar token, also in the absence of accretion an initially stable disk will become unstable if it is able to cool rapidly enough compared to the viscous time scale, see \citet{Gammie01}. 

For accretion disks that are continuously fed by infaling gas from an extended envelope, in our case from the gas that is further out in the halo, the Toomre criterion can be extended. \citet{Kratter2010}, for example argue that two dimensionless parameters can be used to address the stability of mass-loaded disks. The primary parameter, 
\begin{equation}
\xi = \dfrac{\dot{M}_{\rm in}}{\dot{M}_{\rm disk}} = \dfrac{c^3_{\rm s,halo}/G}{c^3_{\rm s,disk}/G} = \dfrac{T^{3/2}_{\rm halo}}{T^{3/2}_{\rm disk}}\;,
\label{eq:xi}
\end{equation}
compares the effective sound speed, or equivalently the effective temperature, in the halo with the corresponding value in the disk. Effective means here that the normal thermal sound speed could be increased by the presence of microturbulence or by the presence of magnetic fields, see Equation (\ref{eq:effective-sound-speed}). The effective temperature is then simply $T_{\rm eff} = \rho c^2_{\rm s, eff} / k $. Equation (\ref{eq:xi}) makes use of the fact that in isothermal collapse models the accretion rate scales as $c_{\rm s}^3/G$, see Equation~(\ref{eq:accretion-rate}). This relation can also be used to characterize the mass flow through  disks in steady state \citep{Kratter2016}. The interpretation of  (\ref{eq:xi}) is simple, if the mass load onto the disk from the halo (numerator) exceeds the accretion from the disk onto the central protostar (denomintor),  then the disk becomes Toomre unstable and will fragment. With other words, instability is likely to occur if the halo gas is effectively hotter than the disk material, say because of the presence of strong microturbulence in the halo or because the disk is so dense that additional cooling processes can work there, see the reviews \citet{Glover2005, Glover2013}. Numerical simulations indicate critical values of $\xi \gtrsim 3$ for fragmentation to set in \citep{Offner2010}.

A secondary parameter, 
\begin{equation}
\Gamma = \dfrac{\dot{M}_{\rm in}}{M_{\rm disk} \Omega} \;,
\label{eq:gamma}
\end{equation}
compares the infall timescale $M_{\rm disk}/\dot{M}_{\rm in}$ to the orbital timescale $\Omega$ in the disk. If $\Gamma \ll 1$ then the disk radius is governed by the angular momentum of the infalling material and not by viscous spreading. Roughly speaking the parameter $\Gamma$ compares the relative strength of gravity versus rotation in the halo.

If these concepts are applied to the accretion disks around the first stars, it is found that fragmentation is a widespread phenomenon and that first stars typically form as members of multiple systems with a wide range of masses (see section~\ref{subsec:fragmentation}).

\section{Environment of first star formation}
\label{sec:environment}

Before we discuss further details of protostellar collapse and the evolution of the accretion disks around the first stars let us briefly review the large-scale environment and the thermodynamic properties of the primordial gas.

\subsection{Cosmological context}
\label{subsec:context}

The formation of the first stars in the Universe occurs in systems which have reached sufficiently large masses so that gas cooling becomes important and baryons can go into run-away collapse within a dark matter halo. In the current $\Lambda$CDM paradigm \citep{Planck2016}, gravitationally bound objects form in a hierarchical fashion with smaller objects decoupling earlier from the Hubble flow of cosmic expansion (see also chapter~\ref{bjoern_chapter}). Calculations of the growth of density perturbations in an expanding Universe \citep{Barkana2001} show that the corresponding Jeans mass scales with redshift $z$ as well as cosmological matter and baryon density parameters, $\Omega_{\rm m}$ and $\Omega_{\rm b}$, as
\begin{equation}
M_{\rm J} \approx 5\times10^3 {\rm M}_{\odot} \left( \dfrac{\Omega_{\rm m}h^2}{0.14} \right)^{-1/2} \left( \dfrac{\Omega_{\rm b}h^2}{0.022} \right)^{-3/5} \left( \dfrac{z+1}{10} \right)^{3/2}\;.
\label{eq:Jeans-halo}
\end{equation}
Here $h$ is the value of the Hubble parameter in units of $100\,$km$\,$s$^{-1}\,$Mpc$^{-1}$, and the parameters are normalized to the 2015 Planck data. However, the criterion for gas to be bound is not sufficient. It also needs to be able to cool for gravitational collapse to set in and lead to first star formation. Considering the various cooling channels of primordial gas described in chapter~\ref{chapter_chemistry}, see also \citet{Glover2005,Glover2013}, this leads to another critical mass scale of
\begin{equation}
M_{\rm cool} \approx 6\times10^5 {\rm M}_{\odot} \, h^{-1} \Omega_{\rm m}^{-1/2} \left( \dfrac{\mu}{1.22} \right)^{-3/2} \left( \dfrac{z+1}{10} \right)^{3/2}\;,
\label{eq:cool-halo}
\end{equation}
where again the mean molecular weight $\mu$ of primordial gas enters the equation. 

We note that below a redshift of $z\approx 40$ the critical mass from cooling $M_{\rm cool}$ becomes larger than $M_{\rm J}$ and so we expect a significant number of halos  that are not able to form stars, because their gas is not able to cool sufficiently fast (that is within a Hubble time). We also note that there are several processes that can increase the critical mass for collapse even further. For example the relative streaming velocity between baryons and dark matter \citep{Tseliakhovich2010, Fialkov14} is able to add a turbulent contribution to the effective sound speed in Equation (\ref{eq:Jeans-mass}) and so can delay collapse (section~\ref{subsec:streaming}). Similar holds for the slow build up of background radiation in the Lyman-Werner band with photon energies from 11.2$\,$eV to 13.6$\,$eV or cosmic reionization at later times, where the additional heat input or change of chemical composition may again delay the gravitational collapse of gas in halos with masses that exceed $M_{\rm cool}$. Clearly, this is most extreme in the case of halos that lie in close vicinity to sites of first star formation, where this external feedback may lead to the formation of supermassive stars \citep{Agarwal12, Agarwal2016}, which in turn could be the progenitors of the supermassive black holes that we observe in the centers of some galaxies at high redshift \citep{Mortlock2011,Wu15}. The latter will be discussed in more detail in chapter~\ref{dcbh}.

Altogether,  first star formation is likely to start at a redshift $z \gtrsim 30$ in rare high-sigma fluctuations, then rapidly becomes possible in more and more halos, and reaches a peak rate at redshifts $z \sim 20 - 15$. Although the overall cosmic star formation rate continues to increase \citep{Madau2014}, the rate at which  metal-free stars form declines again. Regions in the Universe that are not enriched by supernova ejecta from massive stars become increasingly rare. This is the transition to so-called Population II star formation (see section~\ref{subsec:second-gen-stars}). When exactly the formation of genuine Population III stars ends is difficult to assess, and different models and numerical simulations give vastly different answers. It is conceivable that some very low-density regions in voids in the cosmic web have not yet been polluted by metals and so the formation of genuine Population III stars is still possible. 

\begin{figure}[ht]
\begin{center}
\includegraphics[width=0.90\textwidth]{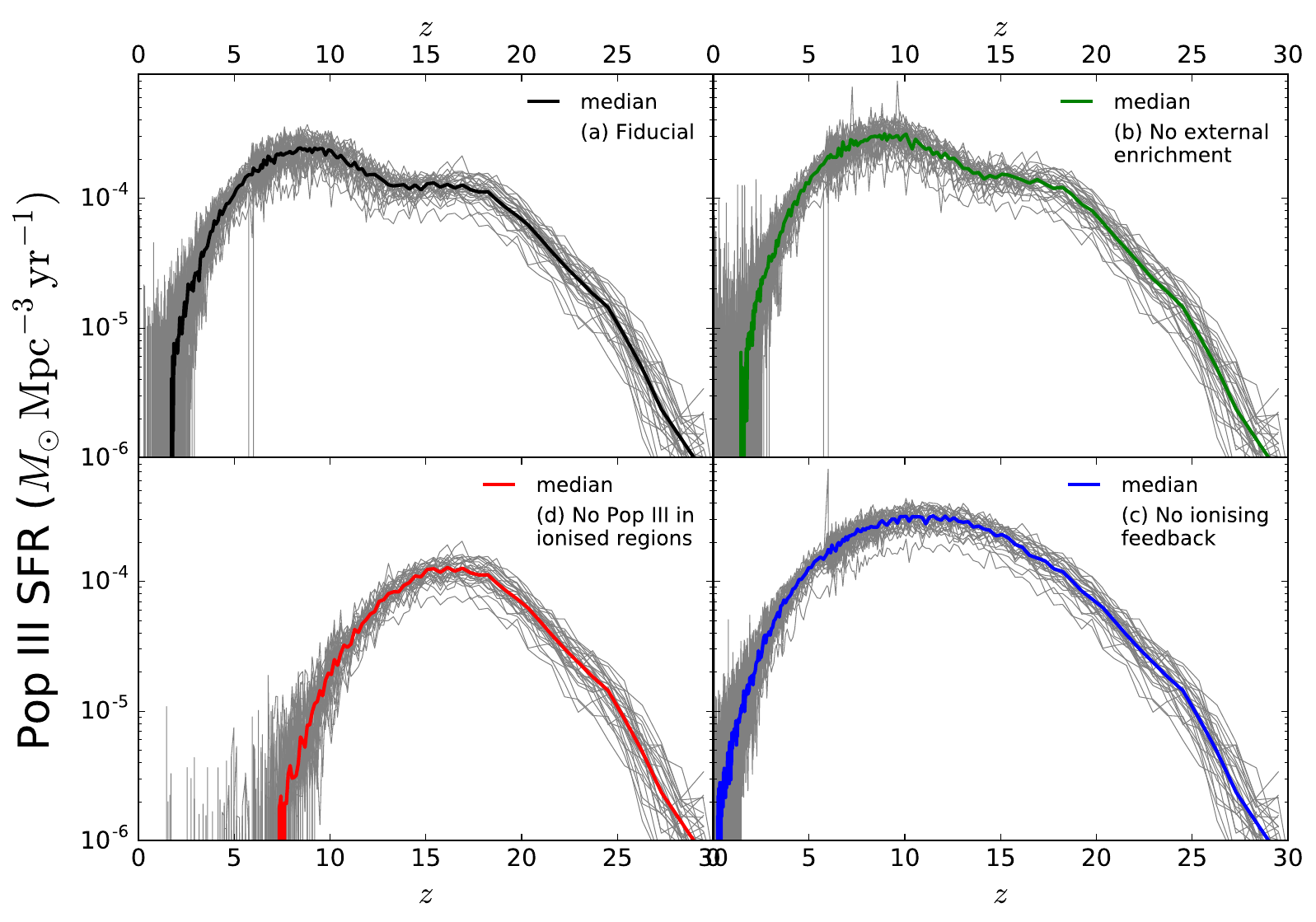}
\end{center}
\caption{
Comoving Pop. III star formation rate densities in units of M$_\odot\,$Mpc$^{-3}\,$yr$^{-1}$ and their medians. The values are calculated from a suite of 30 Local Group simulations  named Caterpillar,see {Griffen16} using four different assumptions about stellar feedback and metal enrichment. Full details are provided in the main text. Adopted from \citet{Magg2017}.
}
\label{fig:Magg}
\end{figure}

A typical example of the formation rate of metal-free stars predicted for regions similar to the Local Group is shown in Figure \ref{fig:Magg}. It shows comoving Pop. III star formation rate densities from \citet{Magg2017}. The semi-analytic calculations are based on a suite of 30 Local Group simulations  \citep{Griffen16} using four different assumptions about stellar feedback and metal enrichment.  Panel {\em (a)} gives the fiducial model with radiative feedback and metal enrichment from Pop. III supernovae. The star formation rate has two peaks, one at $z\approx 17$ and one at $z\approx7$. Panel {\em (b)} demonstrates that the metal enrichment from Pop. III supernovae in neighboring halos delays the transition to Pop. II star formation. This model considers only yields from stars within the same halo. Altogether more Pop. III stars are forming. The largest impact comes from switching off ionizing feedback, as illustrated in panel {\em (c)}. This means that individual halos are able to form Pop. III stars for a longer time and this model leads to the highest Pop. III star formation rates at low redshifts. Panel {\em (d)} depicts the model with the most extreme feedback, where star formation is completely shut off inside an ionized region, even in the most massive halos. This model does not allow for Pop. III star formation below a redshift of $z \approx 7$. Overall, Figure \ref{fig:Magg} demonstrates the impact of environmental conditions on the overall rate of Pop. III star formation and provides some estimate for the uncertainties in the current modeling efforts. 

\begin{figure}[ht]
\begin{center}
\includegraphics[width=0.70\textwidth]{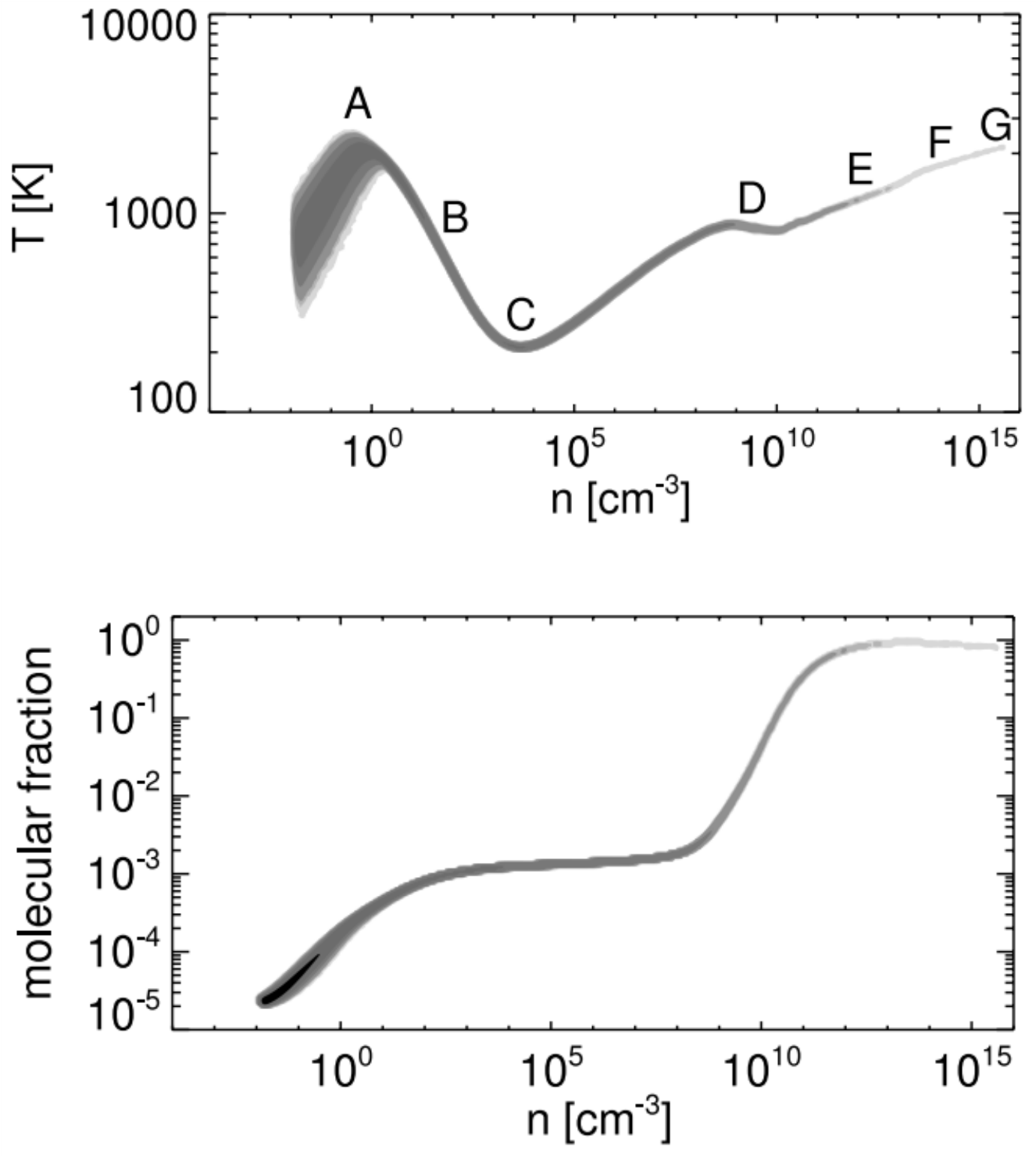}
\end{center}
 \caption{{\em Top panel:} Temperature $T$ of metal-free gas as function of number density $n$. The labels are discussed in the main text. {\em Bottom panel:}  H$_2$ fraction as function $n$. Adopted from \citet{Yoshida2006}.}\label{fig:Yoshida-1}
\end{figure}

\subsection{Thermodynamic behavior of primordial gas}
\label{subsec:thermodynamics}

As discussed above, the ability of gas to collapse and form stars depends on its thermodynamic behavior as  determined by the competition between the available heating and cooling functions. Despite its chemical simplicity several  cooling and heating channels exist for zero-metallicity gas, and become important in different density and temperature regimes.  All cooling processes are related to hydrogen, either in atomic or molecular form. At high temperatures, collisions can populate the first exited electronic states of H which then de-excite by emitting Ly-$\alpha$ photons. This process is most efficient around temperatures of $\sim 10^4\,$K. To reach lower temperatures, molecular hydrogen is needed. Since H$_2$ is a symmetric molecule, it has no permanent dipole moment, and so only quadrupole or higher order transitions are permitted. The corresponding rates are relatively small. In addition, H$_2$ exists in two states with either parallel nuclear spins (para-hydrogen) or anti-parallel spins (ortho-hydrogen). Transitions between para- and ortho-states are forbidden, so that the lowest allowed transition occurs between the $J=2$ and $J=0$ rotational levels in the vibrational ground state of para-hydrogen. The transition energy corresponds to a temperature of $\sim 512\,$K. The high-velocity tail in the thermal Maxwell-Boltzmann velocity distribution allows the gas to cool down to about $200\,$K, see \citet{Greif2014}. The temperature can drop even further, if cooling by deuterated hydrogen takes over. The HD molecule has a non-zero dipole moment and it is not separated into para- and ortho-states and so its lowest energy transition from the ground state is between the $J=1$ to $J=0$ rotational levels, corresponding to a temperature of $\sim 128\,$K. In practice, HD cooling becomes significant only in regions with enhanced fractional ionization, for example in very massive or in externally irradiated halos. In most sites of Pop. III star formation this process is not important. For a more detailed account of the cooling and heating processes in primordial gas, see chapter~\ref{chapter_chemistry} or the reviews by  \citet{Glover2005, Glover2013}.

\begin{figure}[h]
\begin{center}
\includegraphics[width=0.70\textwidth]{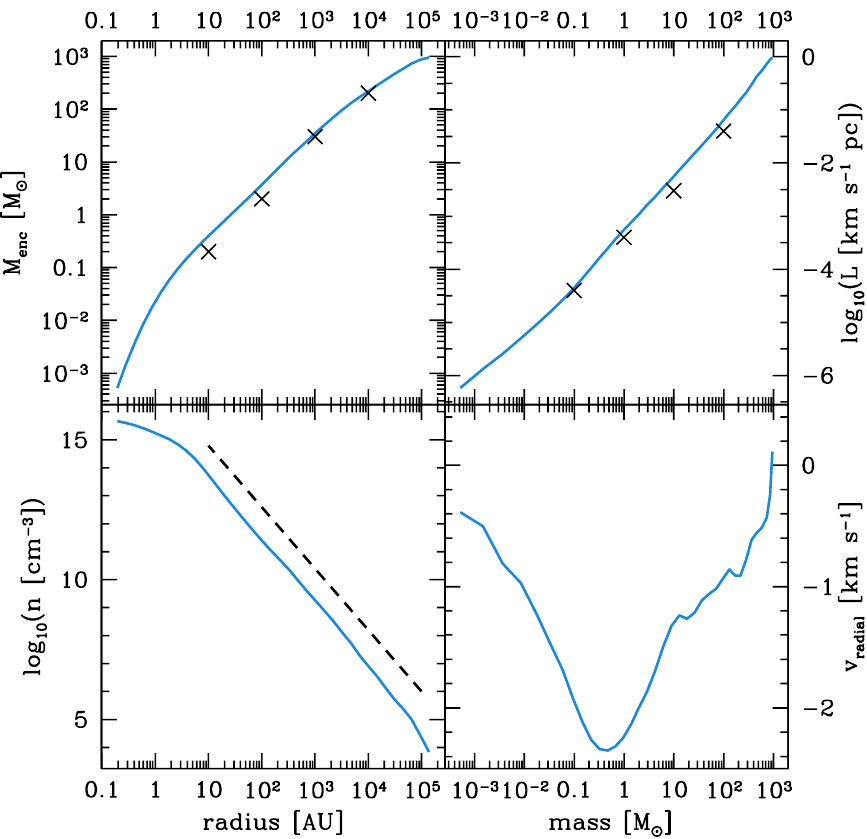}
\end{center}
 \caption{Physical properties of the inner $\sim0.1\,$pc of a star-forming halo, plotted just before the formation of the first star in the center. {\em Upper left:} Enclosed mass as function of distance from the center. {\em Lower left:} Radial number density profile. {\em Upper right:} Specific angular momentum plotted as a function of the enclosed mass. {\em Lower right:} Radial inflow velocity as function of enclosed mass.  The black crosses represent the corresponding values in the simulation presented by \citet{Abel2002}. The dashed line represents the power-law behavior of the density, $n \propto r^{-2.2}$,  typically found in simulations of Pop. III star formation. Figure from \citet{Clark11}.
 }
\label{fig:Clark-1}
\end{figure}

\pagebreak
At low densities, H$_2$ forms mostly by reacting with H$^-$  \citep{Mcdowell1961, Peebles1968}, which itself requires free electrons to form. The reactions are:
\begin{eqnarray}
{\rm H} + {\rm e}^- &\rightarrow& {\rm H}^- + \gamma\;,\\
{\rm H}^- + {\rm H} &\rightarrow& {\rm H}_2 + {\rm e}^-\;.
\end{eqnarray}
There is also a contribution from the interaction with H$^+$ as intermediary molecule \citep{Saslaw1967}:
\begin{eqnarray}
{\rm H} + {\rm H}^+ &\rightarrow& {\rm H}_2^+ +\gamma\;,\\
{\rm H}_2^+ + {\rm H} &\rightarrow& {\rm H}_2 + {\rm H}^+\;.
\end{eqnarray}
These chanels lead to typical molecular fractions of about $10^{-3}$. At high particle densities above $\sim 10^9\,$cm$^{-3}$, the three-body reaction  becomes important \citep{Palla1983}:
\begin{eqnarray}
{\rm H} + {\rm H} + {\rm H} &\rightarrow& {\rm H}_2 + {\rm H} \;.
\end{eqnarray}
As a result all atomic hydrogen is converted into H$_2$ once particle densities of $n \approx 10^{11}\,$cm$^{-3}$ are reached. This is illustrated in the bottom panel of Figure~\ref{fig:Yoshida-1}, which is adopted from \citet{Yoshida2006}. The top panel of this figure shows the corresponding equilibrium temperature $T$ as function of $n$. The labels indicate important phases of the collapse. {\sf (A)} As the gas begins to flow into the potential well of the dark matter halo, it is  compressionally heated to the virial temperature of the system.  Once sufficient molecular hydrogen is formed the gas goes into a run-away cooling phase {\sf (B)}, which brings it down to the minimum temperature of $\sim 200\,$K {\sf (C)}. At this stage of the collapse cold gas can accumulated in the center of the halo. \citet{Bromm2002} call this a 'loitering' phase. When enough gas is accumulated the collapse proceeds and gas slowly heats up again.  Eventually three-body H$_2$ formation becomes important {\sf (D)} and the gas turns fully molecular. The gas remains roughly isothermal over several decades in density. As $n$ increases further, the cloud slowly becomes optically thick {\sf (E)} and the temperature rises again. At densities $n \approx 10^{14}\,$cm$^{-3}$ {\sf (F)} collisional induced emission is an important coolant. When two molecules come close to each other, van der Waals forces can induce a temporary dipole which allows for efficient dipole emission during the interaction time interval , for details see \citet{Omukai1998, Ripamonti2004}. Finally, at temperatures around $2000\,$K collisional dissociation of H$_2$ sets in {\sf (G)}, and the gas becomes molecular again. Altogether, we note that over more than ten orders of magnitude in density, from $n\approx 10^4\,$cm$^{-3}$ to $n\approx 10^{16}\,$cm$^{-3}$, the temperature only rises by a factor of 10 at most. The gas roughly follows a polytropic equation of state, $P \propto n^{\gamma}$ with effective index $\gamma \approx 1.08$ \citep{Omukai2005}. This close to isothermal behavior is essential for allowing the accretion disk in the center of the halo to fragment efficiently (Section~\ref{subsec:fragmentation}). We note that Figure~\ref{fig:Yoshida-1} was calculated in the absence of additional heat sources, the situation may change if radiative feedback from newly formed stars (Section~\ref{subsec:radiation}) or the possible energy input from dark matter annihilation is considered (section~\ref{subsec:DM-annihilation}).

\section{Stellar birth in the halo center}
\label{sec:complexity}
In this section we investigate how star formation progresses once the combination of gravitational collapse and cooling leads to a strong accretion flow into the center of a high-redshift halo, and we discuss the most important physical processes that govern stellar birth on small scales. We argue that the accretion disk that forms around the first object is likely to fragment, which typically results in the formation of a cluster of stars with a wide range of masses rather than the built-up of a single high-mass object. We then speculate about how stellar feedback, the presence of magnetic fields, the potential energy input from dark matter annihilation and the possible existence of large-scale streaming velocities between baryons and dark matter may influence this picture. 

\subsection{Simple spherical collapse}
\label{subsec:1D-collapse}
The most simple model we can construct for describing primordial star formation is the spherically symmetric continuation of the run-away collapse that sets in when a halo exceeds the critical mass for cooling $M_{\rm cool}$ as defined in Equation (\ref{eq:cool-halo}) above. This was modelled extensively in the early 2000's, as discussed by \citet{Abel2002}, \citet{Bromm2002}, and \citet{Yoshida2003}. Although the numerical simulations were fully three-dimensional, the halos considered were relatively round and so assuming spherical symmetry and only considering radial profiles was a very good approximation during the early stages of collapse, see also \citet{Yoshida2006, Yoshida2008} or \citet{Bromm04}.  Figure \ref{fig:Clark-1} adopted from \citet{Clark11} shows enclosed mass and number density as function of radius {\em (left)}, as well as the specific angular momentum and the radial inflow velocity as function of enclosed mass {\em (right)}.  The gas temperature is $\sim 1500\,$K and we can use Equation~(\ref{eq:accretion-rate}) to obtain a good estimate of the measured accretion rate of $\dot{M} \approx 2 \times 10^{-3}\,$M$_{\odot}\,$yr$^{-1}$. 

These early calculations typically stopped when the object in the center reached number densities of $n \approx 10^{16}\,$cm$^{-3}$ or more, because then the computational timestep became prohibitively small.  At this time the protostar has a mass of only $\sim 10^{-3}\,$M$_{\odot}$. At such early stage of evolution the accretion disk has only a small mass and it is strongly sub-Keplerian, meaning that the rotational velocity is smaller than the circular velocity corresponding to centrifugal support, due to the stabilizing influence of large pressure gradients and, if magnetic fields are taken into account, due to the additional support from magnetic pressure and tension. Consequently, this early disk shows no sign of fragmentation at the time these simulations have been stopped. Similar holds for the gas further out in the infaling envelope. The authors of these studies  concluded that the same should be true for the entire protostellar accretion history, and so they argued that all the inflowing mass would end up in one single high-mass star, see also \citet{Tan2004}. Clearly this supposition needed to be tested, in particular, because the idea that primordial stars only form in isolation is in tension with present-day star formation where fragmentation is ubiquitous and young stars are typically found in clusters and aggregates \citet{Lada2003}.

\subsection{Disk fragmentation}
\label{subsec:fragmentation}
In the Universe today, the process of stellar birth is controlled by the intricate interplay between the self-gravity of the star-forming gas and various opposing agents, such as supersonic turbulence, magnetic fields, radiative and mechanical feedback, gas pressure, and cosmic rays. In particular turbulence has been identified to play a key role, see the reviews by  \citet{Maclow2004, Mckee2007, Klessen2016}. On global scales it provides support, while at the same time it can promote local collapse. This process is modified by the thermodynamic response of the gas, which is determined by the balance between various heating and cooling processes. These in turn depend on the chemical composition of the material. Here, clearly, differences between present-day and primordial star formation are expected (see section \ref{subsec:thermodynamics}).

\begin{figure}[th]
\begin{center}
\includegraphics[width=0.85\textwidth]{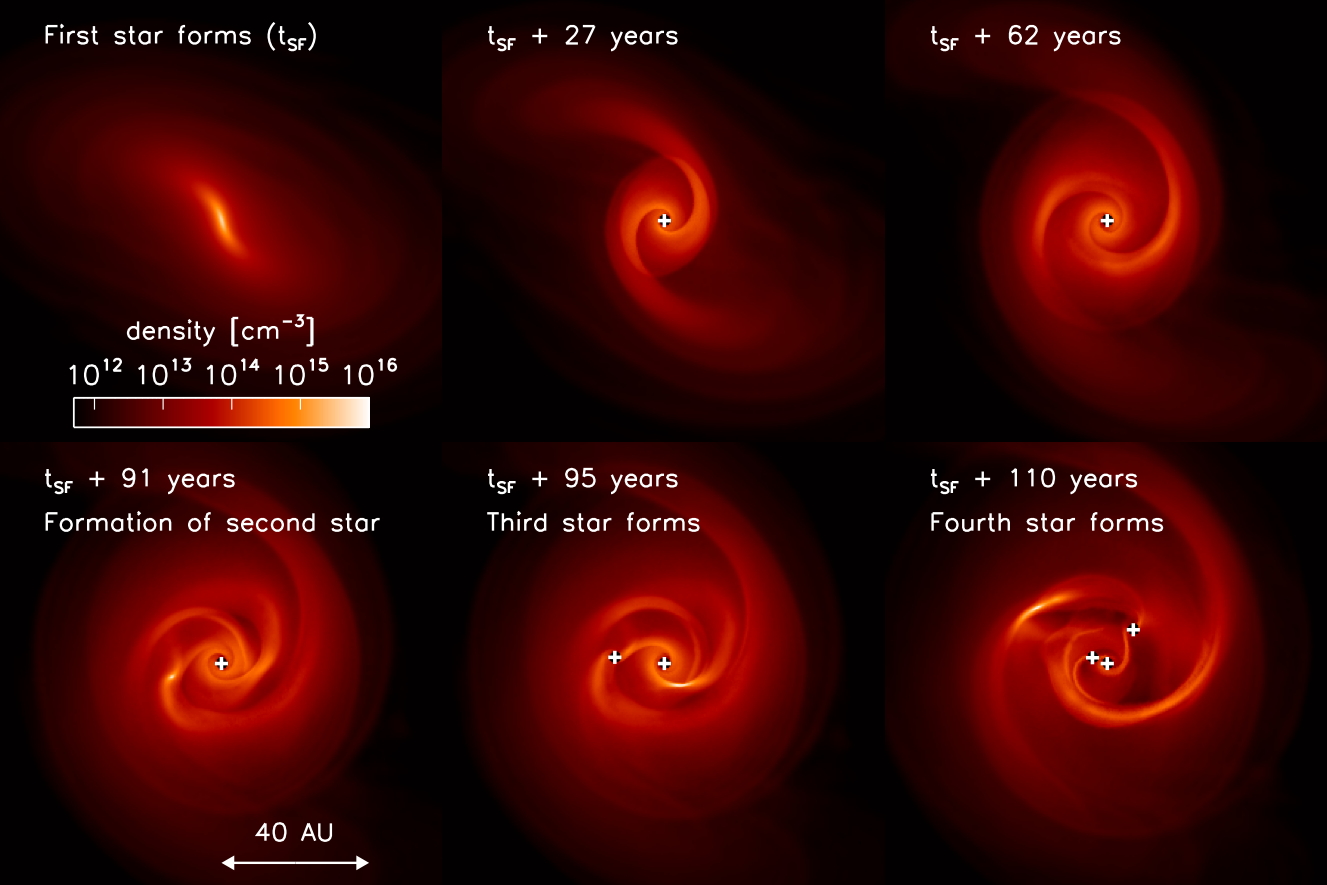}
\end{center}
 \caption{Density evolution in a 120 AU region around the first protostar, showing the build-up of the protostellar disk and its eventual fragmentation. The prominent two-arm spiral structure is caused by the gravitational instability in the disk, and the resulting gravitational torques provide the main source of angular momentum transport that allows disk material to accrete onto the protostar. Eventually, as mass continues to pour onto the disk from the infalling envelope, the disk becomes so unstable that regions in the spiral arms become self-gravitating in their own right. The disk fragments and a multiple system is formed.  Figure adopted from \citet{Clark11}.
 }
\label{fig:Clark-2}
\end{figure}

The introduction of techniques that had been standard repertoire in present-day star formation studies,  in particular the use of sink particles, allowed to continue the numerical simulations beyond the first occurrence of a protostar. In this approach, a contracting high-density region is replaced by a single Lagrangian particle when the resolution limit is reached. The particle inherits the mass as well as the linear and angular momentum of the original region, and it continues to accrete mass that falls in at later times, see \citet{Bate1995,Krumholz2004,Jappsen2005, Federrath2010, Bleuler2014,Howard2014, Sormani2017}. This technique enables calculations that model the build-up and long-term evolution of the accretion disk, and that follow the entire protostar mass growth history.  These studies unambiguously show that primordial accretion disks are highly prone to fragmentation. They indicate that the standard pathway of Pop. III star formation leads to a stellar cluster with a wide distribution of masses rather than the build-up of one single high-mass object. An example is illustrated in Figure \ref{fig:Clark-2}, adopted again from \citet{Clark11}, which shows the evolution of the accretion disks and the build-up of altogether four protostars within only about hundred years after the formation of the first object. 

Disk fragmentation on various spatial and temporal scales is also reported by \citet{Machida2008, Greif2011,Greif12, Dopcke2013,Susa13,Susa14,Hosokawa2016,Stacy2013,Stacy2016,Turk09}. The reason is always the same. Under typical conditions of Pop. III star formation the mass load onto the disk by accretion from the infalling envelope exceeds its capability to transport this material inwards by gravitational or magnetoviscous torques, that is by spiral arms \citep{Binney1987} or by the magnetorotational instability \citep{Balbus1998}. The mass of the disk grows, and it quickly becomes Toomre unstable (section~\ref{subsec:Toomre}). This preferentially occurs at the outer edge of the disk. The instability region moves outwards as the disk grows larger by accretion of higher angular momentum material. And so, fragmentation and the formation of new protostars occurs at larger and larger radii as the evolution progresses, as visible in Figure \ref{fig:Clark-2}.

This has important consequences for the overall accretion history and the resulting mass spectrum. As matter flows through the disk towards the center, it first encounters the Hill volume of secondary protostars further out and preferentially gets swallowed by these objects. Otherwise, this gas would have continued to move inwards and would eventually be accreted by the central object. Clearly, disk fragmentation limits the mass growth of the primary star in the center, and so this  process has been termed 'fragmentation-induced starvation' in the context of present-day star formation \citep{Kratter2006,Peters2010, Girichidis2011, Girichidis2012,Girichidis2012B}. Some of these protostars may get ejected by dynamical encounters with other protostars or fragments while some may move inwards to get accreted by the central object \citep{Clark2011b, Greif12, Smith2012, Stacy2016}. As a result of these highly unpredictable and stochastic events, the mass spectrum of Pop. III stars is expected to cover a wide range of masses, possibly reaching down into the substellar regime. Possible implications of this are discussed in Section \ref{sec:observations}.

We note that some numerical studies indicate that fragmentation may also occur on larger scales in the halo, on scales of the star-forming cloud as a whole \citep{Turk09, Stacy2010, Clark11}. It may happen when the initial turbulence in the halo gas, that is always present at some low subsonic or transsonic level, gets amplified during gravitational collapse and induces density fluctuations that can go into run-away growth in their own right. This is then very similar to the turbulence-driven mode of star formation that is dominant at present days, for reviews see \citet{Maclow2004, Mckee2007,Krumholz2015, Klessen2016}. The importance of this process is enhanced in atomic cooling halos, where cold streams of gas bring dense material towards the center with supersonic velocities (with respect to the cold gas), thereby strongly raising the level of turbulence in the halo \citep{Greif2008, Wise2007, Wise2008}. Similar is also expected in halos that are strongly affected by streaming velocities between baryons and dark matter (section~\ref{subsec:streaming}). It has been suggested that the primordial IMF may be different under more turbulent conditions \citep{Mckee2008, Clark11, Maio2011, Stacy2011}, however, overall the results are not fully conclusive.

\subsection{Radiative feedback}
\label{subsec:radiation}

Since the protostellar Kelvin-Helmholtz contraction time decreases rapidly with increasing stellar mass, massive stars enter the hydrogen burning main sequence while still accreting \citep{Zinnecker2007,Maeder2012}. The properties of the resulting star hereby  depend very strongly on the details of the mass growth history. First stars with accretion rates below values of $\dot{M} \approx 10^{-3}\,$M$_{\odot}$, as we typically expect in low-mass minihalos at high redshift (section~\ref{subsec:1D-collapse}), are compact and very hot at their surface \citep{Hosokawa2009, Hosokawa2010, Hosokawa2012}. These stars  emit copious amounts of ionizing photons \citep{Schaerer2002} that can significantly influence their birth environment, as we discuss below. On the other hand, for rates of $\dot{M} \gtrsim 10^{-2}\,$M$_{\odot}$ stellar evolution calculations suggest that stars remains bloated and fluffy, and because of the large radius the surface temperature is relatively low. Although very luminous, these stars do not emit much ionizing radiation and could be able to maintain this high accretion flow for a very long time. In the right environment this could possibly lead to the formation of supermassive stars  \citep{Hosokawa2013,Haemmerle16, Umeda2016, Woods2017}

\pagebreak
Compact and hot Pop. III stars create HII regions which are likely to break out of the parental halo and affect stellar birth in neighboring halos. Many aspects of this problem have been addressed, for example by \citet{Kitayama2004}, \citet{Whalen2004, Alvarez2006, Abel2007,Yoshida2007,Greif2008,Wise2012,Wise2012b,Jeon2014}. Here we focus on the impact of radiative feedback on the immediate birth environment of the star and on the question of how this influences the fragmentation behavior of the disk and the resulting stellar mass spectrum. 

As before, we can seek guidance from models of present-day star formation. Radiation hydrodynamic simulations in two and three dimensions that take both non-ionizing and ionizing radiation into account \citep{Yorke2002,Krumholz2009, Kuiper2010, Kuiper2011, Peters2010, Peters2011, Commercon2011, Rosen2016} demonstrate that once a protostellar accretion disk has formed, it quickly becomes gravitationally unstable and so material in the disk midplane flows inwards along dense filaments, whereas  radiation escapes through optically thin channels above the disk. Even ionized material can be accreted, if the accretion flow is strong enough \citep{keto2007, Peters2010}. Radiative feedback is thought not to be able to shut off the accretion flow onto massive stars. Instead it is the dynamical evolution of the disk material that controls the mass growth of individual protostars. Accretion onto the central object is shut off by the fragmentation of the disk and the formation of lower-mass companions which intercept inward-moving material as argued in the previous section~\ref{subsec:fragmentation}. This requires three-dimensional simulations, because this fragmentation process is not properly captured in two dimensions.

Due to the lack of metals and dust, protostellar accretion disks around Pop. III stars can cool less efficiently and are much hotter than disks at present days. Similarly, the stellar radiation field couples less efficiently to the  surrounding because the opacities are smaller. It is thus not clear how well the above results can be transferred to the primordial case. Insight can be gained from the two-dimensional radiation hydrodynamic simulations of Pop. III star formation by \citet{Hosokawa11,Hirano2014, Hirano2015}. They follow disk formation and the long-term accretion history of the central object and  find that radiative feedback can indeed stop stellar mass growth and blow away the remaining accretion disk, leading to final stellar masses from a few $10\,$M$_\odot$ up to about $1000\,$M$_\odot$. However, these calculations cannot capture disk fragmentation and the formation of multiple stellar systems. Three-dimensional calculations have been reported by \citet{Stacy2012,Susa13,Susa14,Hosokawa2016}. These studies find widespread fragmentation, again with a wide range of stellar masses down to $\sim 1\,$M$_\odot$. These simulations have their own limitations and shortcomings, either in terms of resolution or in the number of physical processes included. And so, any firm conclusions about the resulting mass spectrum of Pop. III stars in the presence of radiative feedback is premature at this stage. This is clearly one of the frontiers of current research in primordial star formation.

\begin{figure}[th]
\begin{center}
\includegraphics[width=0.50\textwidth]{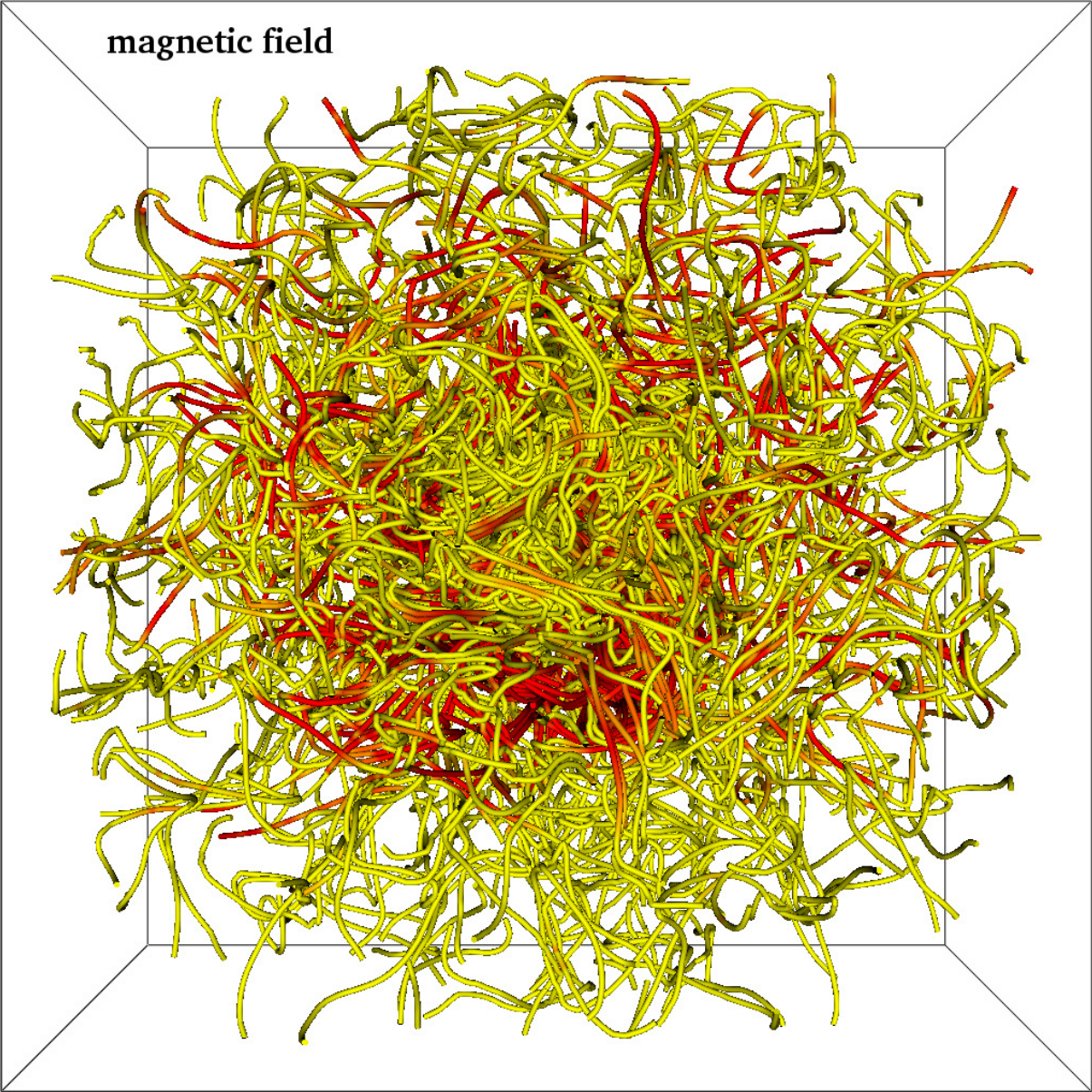}
\caption{Magnetid field configuration in the center of a collapsing halo. The highly tangled field lines illustrate nicely the complexity of the field structure in the presence of a turbulent dynamo. The figure is adopted from \citet{Federrath2011}.
 }
\end{center}
\label{fig:Federrath-1}
\end{figure}

\subsection{Magnetic fields}
\label{subsec:magnetic-fields}

The presence of dynamically important magnetic fields could significantly alter the picture presented so far. We know that the current Universe is highly magnetized on all scales \citep{Beck1996} and that this influences the birth of stars and the evolution of the interstellar medium, for the extreme viewpoint of magnetically mediated star formation, see \citet{Shu1987}. The properties of the magnetic fields observed today are well explained by a combination of small-scale and large-scale dynamo processes \citep{Brandenburg2005}. In contrast, our knowledge of magnetic fields at high redshifts is very sparse. Theoretical models predict that magnetic fields could be produced in various ways, for example via the Biermann battery \citep{Biermann1950}, the Weibel instability \citep{Lazar2009, Medvedev2004}, or thermal plasma fluctuations \citep{Schlickeiser2003}. Other theories place their origin in phase transitions that occur during cosmic inflation \citep{Sigl1997, Grasso2001,Banerjee2003, Widrow2012}. The resulting fields are thought to be orders of magnitudes too weak to have any dynamical impact, and so, magnetohydrodynamic effects have typically been neglected in numerical simulations of primordial star formation  however, see the analytic models by \citet{Pudritz1989,Tan2004,Silk2006}.

This situation has changed, when people started to realize that the small-scale turbulent dynamo is efficiently amplifying even extremely small primordial seed fields to the saturation level, and that this process is very fast, acting on timescales much shorter than the dynamical free-fall time. An analytic treatment is possible in terms of the Kazantsev model \citep{Kazantsev1968, Subramanian1998, Schobera,Schoberb} for low as well as large Prandtl numbers. This theory describes how the twisting, stretching, and folding of field lines in turbulent magnetized flows (see Figure~\ref{fig:Federrath-1}) leads to an exponential growth. The amplification timescale is comparable to the eddy-turnover time on viscous or resistive length scales. This is the kinematic regime. Once  backreactions become important, the growth rate slows down, and saturation is reached within a few large-scale eddy-turnover times \citep{Scheko02, Schekochihin2004, Schober2015}. Depending on the properties of the turbulent flow the magnetic energy density at saturation is thought to lie between 0.1\% and a few 10\% of the kinetic energy density \citep{Federrath2011, Federrath2014}. 

\pagebreak
Magnetic fields with that strength can strongly affect the evolution of protostellar accretion disks. They remove angular momentum from the star-forming gas \citep{Machida2008,Machida2013, Bovino2013NJP, Latif2013a,LatifMag2014}, drive protostellar jets and outflows \citep{Machida2006}, and  reduce the level of fragmentation in the disk \citep{Turk2011, Peters2014}. We expect Pop. III clusters to have fewer members with somewhat higher masses than predicted by purely hydrodynamic  simulations (as discussed in section~\ref{subsec:fragmentation}). However, the details of how magnetic fields influence the overall star-formation process in primordial gas, and how they affect the resulting IMF are not well understood. This remains to be an area of very active research.

\subsection{Relative streaming velocities between baryons and dark matter}
\label{subsec:streaming}

Prior to recombination, baryons are tightly coupled to photons resulting in a standing acoustic wave pattern \citep{Sunyaev1970}. In turn, this leads to oscillations between baryons and dark matter  with relative velocities of about $30\,$km$\,$s$^{-1}$ and coherence lengths of several $10\,$Mpc to $100\,$Mpc \citep{Silk1986} at $z \approx 1000$. After recombination, baryons  are no longer tied to photons, their sound speed drops to $\sim 6\,$km$\,$s$^{-1}$, and the velocity with respect to the dark matter component becomes supersonic with Mach numbers of ${\cal M} \approx 5$. This has been first described by  \citet{Tseliakhovich2010}. In the subsequent cosmic evolution, the relative streaming velocity decays linearly with $z$, and reaches  $\sim 1\,$km$\,$s$^{-1}$ at the onset of first star formation at $z \approx 30$. It is comparable to the virial velocity of the first halos to cool and collapse, and so it has been suggested that this velocity offset strongly influences the star formation process \citep{Tseliakhovich2011, Fialkov2012}. 

Simulations that include streaming velocities indeed suggest that their presence reduces the baryon overdensity in low-mass halos, delays the onset of cooling, and leads to a larger critical mass for collapse to set in \citep{Greif2011, Stacy2011, Maio2011, Naoz2012, Naoz2013, Oleary2012, Latif2014Stream, Schauer17a}. They may also have substantial impact on the resulting $21\,$cm emission \citep{Fialkov2012, McQuinn2012,Visbal2012}. Gas in halos which are located in regions of large streaming velocities will be more turbulent than in more quiescent systems, and so we expect more fragmentation and a bias towards smaller stellar masses \citep{Clark2008}. However, this process has not yet been modeled with sufficient resolution, and so a reliable prediction of the IMF in regions with large streaming velocities is still outstanding.

\subsection{Potential impact of dark matter annihilation}
\label{subsec:DM-annihilation}

Despite its importance for cosmic evolution and structure formation, the true physical nature of dark matter is still unknown. Many models introduce a new class of  weakly interacting massive particles, so called WIMPs, as they naturally occur in supersymmetry theories. The lightest supersymmetric particle is expected to be stable and to have properties that agree with the phenomenological requirements on dark matter  \citep{Bertone2005}. Specifically, they are self-annihilating Majorana particles, that is they are their own antiparticles, and they interact only weakly with baryons.

\pagebreak
In most environments the dark matter density is much too low for this to be significant. However, it has been suggested to be different in star-forming halos in the early Universe. Here the collapse of baryons may drag along dark matter particles. This process is called 'adiabatic contraction' \citep{Blumenthal1986}, and it can lead to a density increase of several orders of magnitude in the very center of the halo. As the annihilation rate scales quadratically with the density the corresponding energy input and ionization rate may become large enough to influence the dynamics of the gas. \citet{Spolyar2008,Freese2009} realize that this process could potentially overcome the cooling provided by H$_2$. They speculate that this could halt further collapse and lead to the formation of a dark star, an object of a few AU in size that is powered by dark matter  annihilation rather than by nuclear fusion. If dark matter particles also scatter weakly on baryons this could lead to a structure that is stable for a long time. These dark stars are thought to be much larger than normal Pop. III stars, have lower surface temperatures, and are more luminous \citep{Freese2008, Iocco2008A,Iocco2008B, Yoon2008,Hirano2011}. 

There are several problems with this scenario. First, it is not clear whether collapse stalls once the energy input from dark matter becomes comparable to the cooling rate. \citet{Ripamonti2010}, for example argue that this may not be the case because the larger heating rate catalyzes further formation of H$_2$ and is compensated by the corresponding larger cooling rate. Second, the inherent assumption of perfect alignment between dark matter cusp and gas collapse is most likely violated in realistic star formation conditions. There is always some degree of anisotropy in the halo and there is always some level of turbulence in the infalling gas that make it highly unlikely for the dark matter cusp and the collapsing gas to overlap perfectly. This problem was studied by \citet{Stacy2012, Stacy2014}, who indeed found that non-axisymmetric perturbations lead to a separation between dark matter cusp and collapsing gas, rendering the annihilation energy input insignificant for dark star formation. Similar was reported by \citet{Smith2012}, who also found no evidence for a dark stellar phase in their calculations, but instead formed more normal Pop. III stars. These authors, however, also reported that dark matter annihilation was able to influence the dynamics of the accretion disk and that the energy input associated with this process lead to a suppression of disk fragmentation. They concluded that dark matter annihilation may reduce the multiplicity of metal-free stars and increase their overall mass. But, as before, the existing simulations are still too premature to draw firm conclusions about the Pop. III IMF.

 \subsection{Second generation star formation}
 \label{subsec:second-gen-stars}
 
Second generation stars, sometimes termed Population II.5 stars, have formed from material that has been enriched from the debris of the first stars. Unlike the very first stars, for which we have no direct detections yet (section~\ref{subsec:indirect-high-z}), members of the second generation have already been found in surveys looking for extremely metal-poor stars in our Milky Way and neighboring satellite galaxies (section~\ref{subsec:direct-local}). There are two competing models for their origin based on different low-metallicity cooling channels that determine the ability of the gas to  collapse and fragment, and so set the stellar mass spectrum. 

Unlike purely primordial gas, where cooling is solely provided by hydrogen in various forms (section~\ref{subsec:thermodynamics}), metal-enriched gas has access to a wide range of different cooling processes. Atomic fine-structure lines from alpha elements such as carbon or oxygen can provide sufficient cooling at number densities around $10^4\,$cm$^{-3}$ above a critical metallicity of about $10^{-3}$ of the solar value \citep{Smith2009}. Details of the element abundance in the Sun are discussed by \citet{Asplund2009,Caffau2011}. Up to $n\approx 10^2\,$cm$^{-3}$ and metallicities of $\sim 10^{-2}$ solar the cooling rate is dominated by molecular hydrogen \citep{Jappsen2007, Jappsen2009, Glover2014}. This can be estimated by calculating the amount of heavy elements required to produce a cooling rate equal to the rate of adiabatic compression heating for given halo properties \citet{Bromm03, Santoro2006}. \citet{Frebel2009} combined the available abundance measures and introduced a transition criterion, with low-mass stars forming only above a certain threshold. This proposition is challenged by the discovery of SDSS J1029151+172927, a truly primitive star in the constellation of Leo, that  falls below this value \citep{Caffau2011Nat}. The star has elemental abundances in the range $10^{-5}$ to $10^{-4}$ of the solar value for all of the elements measured in its spectrum, setting it apart from other extremely metal-poor stars which typically have enhanced CNO abundances despite being very iron poor. 

Another line of reasoning considers dust cooling as the primary agent for fragmentation, which leads to a lower critical metallicity in the range $10^{-5}$ to $10^{-6}$ \citep{Omukai2005, Schneider2006, Schneider2012,Chiaki2013, Chiaki2013B}. The existence of SDSSJ1029151+172927 provides evidence for the validity of the dust-induced fragmentation model \citep{Schneider2012,Klessen2012, Bovino2016}. The analytical models, based on comparing the strength of various cooling and heating processes in a metal-poor environment, are supported by numerical simulations \citep{Tsuribe2006,Tsuribe2008, Clark2008, Dopcke2013, Chiaki2014, Chiaki2016}. These suggest that indeed dust is responsible for a transition to a star formation mode similar to the one observed at present days. This transition occurs at a metallicity of about $10^{-5}$ solar and leads to an IMF peaking below  $1\,$M$_{\odot}$ with  a functional form similar to the one inferred for the solar neighborhood \citep{Kroupa2002, Chabrier2003}.

\section{Observational constraints}
\label{sec:observations}
Determining the properties of the first population of stars is a difficult task. Most of our current knowledge of Pop. III stars comes from  theoretical model building and numerical simulations. To a large degree, this is due to the fact that, unlike in present-day star formation, stringent observational constraints are rare and extremely difficult to obtain. Here we provide a brief overview of possible indirect constraints from the primordial as well as the present-day Universe, and we speculate about possible detections of genuine low-mass Pop. III stars in Galactic archeological surveys. 

\subsection{Indirect constraints from high redshift observations}
\label{subsec:indirect-high-z}

Unfortunately, Pop. III stars are very hard to see directly in the high-redshift Universe. They are too faint \citep{Schaerer2002} to be within reach of the next generation of telescopes. This holds in space  for the James Webb Space Telescope \citep{Gardner2006}, Euclid \citep{Laureijs2011}, or the Wide-Field Infrared Survey Telescope \citep{Spergel2015}, and it is true for ground-based observatories such as the European Extremely Large Telescope \citep{Tamai2014} or  the Thirty Meter Telescope \citep{Skidmore2015}. Constraints on the IMF from high-$z$ observations therefore are indirect at best. For example we can search for the supernova explosions that accompany the end of massive Pop. III stars. Numerical simulations indicate that the light-curves of Pop. III pair-instability supernovae and even less energetic core collapse supernovae will be within easy reach of JWST or E-ELT   out to very high redshifts \citep{Hummel2012, Kasen2011, Pan2012, Whalen2013,Whalen2013b}. However, because of their narrow field of view, and for the space missions also because of the long slewing time, these telescopes are not suited for large area surveys and will have difficulties to find these supernovae. Lists of suitable  target candidates are needed. Here, gravitational lensing might help and bring some of the supernova events above the sensitivity limit of large-area surveys such as planned with LSST,  WFIRST or Euclid , see also \citep[][Rydberg et al., in prep.]{Oguri2010}. Once a candidate is found,  JWST or E-ELT can be used for detailed follow-up observations. The detection rates in large surveys furthermore  can contribute to distinguish between different cosmological models \citep{Magg2016}. 

Studying the high-redshift Universe will only allow us to address the high mass end of the IMF. Information about the low-mass Pop. III stars is completely inaccessible via this route. Similar is true for potential detections of gravitational wave emission from merging Pop. III black hole binaries. The recent discoveries of gravitational waves from three systems with masses of $62\,$M$_\odot$, $21\,$M$_{\odot}$, and $49\,$M$_\odot$ \citep{Abbott2016, Abbott2016A, LVC-GW170104} has demonstrated that this might indeed be possible. Mergers of very massive Pop. III remnants could contribute to the statistical background \citep{Kohei, Sasaki2016} or may even be detected directly \citep{Belczynski2017, Hartwig2016b, Nakama2017}. More details on detecting mergers of black hole binaries from Pop. III stars via gravitational wave emission will be given in chapter~\ref{gws_ch3}. We should also mention  studies by \citet{Kulkarni2013, Kulkarni2014} who investigate the impact of the chemical enrichment from Pop. III stars on damped Lyman-$\alpha$ systems and on cosmic reionization, and find that this could be a sensitive probe of the primordial IMF at high redshift.  

At the extreme end of the mass spectrum, we note that the discovery of very massive quasars with several $10^9\,$M$_\odot$ at redshifts of $z \approx 6$ or above \citep{Mortlock2011, Wu15} provides support for the existence of supermassive stars in the early Universe.  We note that stellar evolution calculations with very high accretion rates suggest that stars could be stable up to several $10^5\,$M$_\odot$ before the general relativistic instability \citep{Iben1963, Chandra1964} leads to the collapse into a black hole of the same mass \citep{Hosokawa2013, Umeda2016, Woods2017}. These could be the seeds for the observed very massive quasars when able to maintain high accretion rates close to the Eddington limit down to $z \approx 6$. The formation of such objects via gravitational collapse will be described in chapter~\ref{dcbh}, and their subsequent evolution in chapter~\ref{super}.

\subsection{Indirect constraints from observations in the Local Group}
\label{subsec:indirect-local}

More stringent limits on the IMF are likely to come from the study of nearby stars in the Local Group. Current Galactic archeological surveys \citep{Beers2005,Caffau2013, Frebel2010} in the halo and bulge of our Milky Way or the analysis of stars in nearby satellite dwarf galaxies \citep{Koch2013,Salvadori2015,Kirby2015,Roederer2016,Ji2016A, Ji2016B}  can contribute to our knowledge of primordial stars in several ways. There is a large body of work that uses the abundance pattern determined in extremely metal-poor stars  to infer the properties of the progenitor stars which provided the enrichment \citep{Heger2002, Heger2010}. Assuming that the oldest and most metal-poor stars in the Galaxy have been supplied with heavy elements by only one or at most two supernova explosions \citep{Chan2017}, it turns out that the measured relative abundances of heavy elements are most consistent with core collapse supernovae from Pop. III stars in the mass range $20 - 40\,$M$_\odot$ \citep{Aoki2014, Bonifacio2015, Caffau2012, Cooke2014, Frebel2005, Iwamoto2005, Joggerst2009, Joggerst2010, Keller2014, Lai2008, Norris2013}. Together with the fact that no genuine signatures of pair-instability supernovae from massive stars in the range of $\sim 130$ to $250\,${M$_\odot$} have been found, this places strong constraints on the high-mass end of the primordial IMF. 

\subsection{Possible direct detections of low-mass Pop. III stars}
\label{subsec:direct-local}

The theory of stellar evolution  \citep{Kippenhahn2012} tells us that any low-mass Pop. III star with $0.8\,${M$_\odot$} or smaller must have survived until the present day. If these stars ever existed, then there is a chance to directly detect some of them in current or future stellar archeological surveys.  Even non-detections allow us to put stringent limits on the low-mass end of the Pop. III IMF. For example \citet{Hartwig2015MNRAS} estimate the expected numbers of low-mass Pop. III stars in the Galactic halo based on semi-analytic models of the early star formation history in Milky Way-like halos. They conclude that if no such object is found in surveys with sample sizes of 4 million stars then we can exclude the existence of low-mass Pop. III stars with masses below $0.8\,${M$_\odot$} with a confidence level of 68\%. This limit may soon be reached. 

On similar grounds, \citet{Salvadori2007, Salvadori2010} develop a detailed model of the metallicity distribution function of metal-poor stars in the Galactic halo and suggest that Pop. III stars should be more massive than $0.9\,${M$_\odot$} when matching their predictions with the observational data. On the other hand, \citet{Tumlinson2006, Tumlinson2010} argues that the current abundance measurements are not really sufficient to distinguish between different models of the primordial IMF based on studying the chemical evolution during the early build-up of the Galaxy. However, he also suggests that characteristic masses of a few $10\,${M$_\odot$} provide a better fit to the available data than masses of $100\,${M$_\odot$} or above, consistent with the supernova yields mentioned above.  

\citet{Mapelli2006} focus on the density of Galactic intermediate-mass black holes. They take them as the relics of higher-mass Pop. III stars and derive an upper limit by comparing their model to the non-detection of ultra-luminous x-ray sources in the Galaxy. This constraint, however, does not affect the low-mass end of the primordial IMF. The approach has been extended by \citet{Bennassuti2014} who also conclude that the Galactic metallicity function and the abundances measured in extremely metal-poor stars in the halo are inconsistent with pair-instability supernovae. Their study indicates that the early enrichment of the Galaxy by Pop. III stars was dominated by core collapse supernovae, and they suggest an IMF that goes all the way down to $0.1\,${M$_\odot$}.  Altogether, the prospects of finding surviving low mass first stars in our immediate neighborhood are highly exciting, but at present no clear conclusions are possible.

\pagebreak
\section{Summary}
\label{sec:summary}

In this chapter of the book we aimed at providing a concise overview of our current understanding of the physical processes that govern stellar birth in the primordial Universe. As the first stars are too faint to be directly detectable at high redshift, the available observational data give little input and provide very indirect constraints at best. And so, most of the current progress in this rapidly evolving field of astrophysical research has been driven by numerical simulations and theoretical model building. 

While early models predicted that the first stars formed in isolation with only one single high-mass star in the center of a halo, the community now acknowledges that the birth of stars in truly metal-free gas is subject to the same complexity and stochasticity that is well known in present-day star formation. Probably the most important development in the past decade is the realization that the accretion disks that build up around the first stars are highly susceptible to fragmentation. The natural formation pathway therefore leads to Population III stars that are members of multiple stellar systems and clusters. This has important consequences for the stellar mass spectrum. Numerical simulations of first star formation that are able to resolve the evolution of the accretion disk and the mass growth of individual protostars with high resolution predict a wide range of masses, with a relatively flat distribution that spans the substellar regime up to several hundred solar masses, and with the most likely values being around a few tens of solar masses. This is different from the IMF observed at present days, which shows a clear peak at $\sim 0.3\,$M$_\odot$ and a power-law fall-off towards higher masses. So, one of the key questions is when and where did the transition between the primordial and the current mode of star formation occur, and what is its primary physical cause. 

A convincing answer to this question is still outstanding. Besides the lack of a good observational base, the primary reason for this is that all existing calculations have severe shortcomings. And so still very little is actually known about the mass spectrum of Pop. III stars. The current numerical simulations either lack sufficient resolution, or they cover only a short fraction of the overall star formation timescale.  Most of them ignore important physical processes, such as protostellar feedback or magnetic fields. It may also be that more speculative phenomena such as heating from dark matter annihilation or the relative streaming velocity between baryons and dark matter are important for the star formation process in the center of a halo. The theoretical and numerical treatment of these processes is still in its infancy. Furthermore, some simulations are two-dimensional, whereas a full three-dimensional approach is required to properly capture gas fragmentation or to fully assess the dynamical impact of stellar radiation on the infaling gas. So, correctly modelling stellar birth in high-redshift halos with high physical fidelity with strong predictive power is still an unsolved problem. It is a true frontier of computational astrophysics. 

It is probably fair to say that most researchers in the field agree that the genuine Pop. III stars have formed in binary or higher-order multiple stellar systems and that they most likely had a wide range of masses. Whether the mass scale reached down to low-mass stars or even the sub-stellar regime depends on physical processes that we only now start to consider properly, such as radiative feedback, the level of turbulence in halo gas and accretion disk, properties of the dark matter, or the larger scale environment of the star forming halo.  Since low-mass Pop. III stars must have survived until the present days, they should be detectable in stellar archeological surveys in the Milky Way and neighboring satellite galaxies, and so there are good chances that we soon have stringent constraints on the low-mass end of the Pop. III IMF. The abundance patterns we observe in extremely metal-poor stars nearby are all consistent with enrichment by core collapse supernovae with progenitor masses of a few tens of solar masses. This is in agreement with the most likely stellar masses from fragmentation calculations of primordial accretion disks. The fact that no signatures of enrichment by pair-instability supernovae are seen, suggests that Pop. III stars with masses above $100\,$M$_\odot$ must have been extremely rare. How far up the mass range of primordial stars extends is not clear yet. Theoretical limits from including general relativistic instabilities in the stellar structure equations predict an upper mass limit of several $10^5\,$M$_\odot$. Indeed, such stars could be the seeds for the extremely massive quasars with billions of solar masses we observe at redshifts of $z \approx 6$ and higher. However, the density of these objects is only a few per Gpc$^3$, and so our statistical base is scarce.  

In this chapter, we have focused on what could be called the standard pathway to Pop. III star formation. We have considered cooling and collapse of truly pristine gas that has not been affected by any form of feedback from earlier episodes of star formation either within the same halo or in neighboring halos. The temperature and composition of the material has not been altered by strong Lyman-Werner or ionizing fluxes, and there was no enrichment from the supernovae of previous stars. These effects are discussed in detail in other chapters of this book. Partiicularly, the next chapter{\ref{dcbh} will discuss the formation of very massive black holes via direct collapse, in situations where the cooling is suppressed by radiation backgrounds. An alternative formation channel via collisions in dense stellar clusters will be presented in chapter~\ref{katz}, and the evolution of supermassive stars is discussed in chapter~\ref{super}. Predictions for gravitational wave emission from black hole binary mergers, including remnants from Pop. III stars, are summarized in chapter~\ref{gws_ch3}.

\section{Acknowledgements}
\label{sec:acknowledgements}

Writing this text would not have been possible without the input and the support of many friends and colleagues in Heidelberg and around the world. I would also like to thank Christoph Federrath, Simon Glover, and Naoki Yoshida for carefully reading this manuscript and for providing very helpful feedback. 
%
%
%
I also acknowledge financial support by the European Research Council via the ERC Advanced Grant {\em STARLIGHT: Formation of the First Stars} (project number 339177), and by the Deutsche Forschungsgemeinschaft (DFG) in the Collaborative Research Center (SFB 881) {\em The Milky Way System} (subprojects B1, B2, and B8) and in the Priority Program SPP 1573 {\em Physics of the Interstellar Medium} (grant numbers KL 1358/18.1, KL 1358/19.2).

%% file: klessen.bbl
\begin{thebibliography}{259}
\newcommand{\enquote}[1]{#1}
\providecommand{\natexlab}[1]{#1}
\providecommand{\url}[1]{\texttt{#1}}
\providecommand{\urlprefix}{URL }
\providecommand{\eprint}{eprint }
\expandafter\ifx\csname urlstyle\endcsname\relax
  \providecommand{\doi}[1]{doi:\discretionary{}{}{}#1}\else
  \providecommand{\doi}{doi:\discretionary{}{}{}\begingroup
  \urlstyle{rm}\Url}\fi

\bibitem[{{Abbott} \emph{et~al.}(2016{\natexlab{a}}){Abbott}, {Abbott},
  {Abbott}, {Abernathy}, {Acernese}, {Ackley}, {Adams}, {Adams}, {Addesso},
  {Adhikari} and et~al.}]{Abbott2016}
{Abbott}, B.~P., {Abbott}, R., {Abbott}, T.~D., {Abernathy}, M.~R., {Acernese},
  F., {Ackley}, K., {Adams}, C., {Adams}, T., {Addesso}, P., {Adhikari}, R.~X.,
   and et~al. (2016{\natexlab{a}}). \enquote{{GW150914: The Advanced LIGO
  Detectors in the Era of First Discoveries},} \emph{Physical Review Letters}
  \textbf{116}, 13, 131103.

\bibitem[{{Abbott} \emph{et~al.}(2016{\natexlab{b}}){Abbott}, {Abbott},
  {Abbott}, {Abernathy}, {Acernese}, {Ackley}, {Adams}, {Adams}, {Addesso},
  {Adhikari} and et~al.}]{Abbott2016A}
{Abbott}, B.~P., {Abbott}, R., {Abbott}, T.~D., {Abernathy}, M.~R., {Acernese},
  F., {Ackley}, K., {Adams}, C., {Adams}, T., {Addesso}, P., {Adhikari}, R.~X.,
   and et~al. (2016{\natexlab{b}}). \enquote{{Observation of Gravitational
  Waves from a Binary Black Hole Merger},} \emph{Physical Review Letters}
  \textbf{116}, 6, 061102.

\bibitem[{{Abbott} \emph{et~al.}(2017){Abbott}, {Abbott}, {Abbott}, {Acernese},
  {Ackley}, {Adams}, {Adams}, {Addesso}, {Adhikari}, {Adya} and
  et~al.}]{LVC-GW170104}
{Abbott}, B.~P., {Abbott}, R., {Abbott}, T.~D., {Acernese}, F., {Ackley}, K.,
  {Adams}, C., {Adams}, T., {Addesso}, P., {Adhikari}, R.~X., {Adya}, V.~B.,
  and et~al. (2017). \enquote{{GW170104: Observation of a 50-Solar-Mass Binary
  Black Hole Coalescence at Redshift 0.2},} \emph{Physical Review Letters}
  \textbf{118}, 22, 221101.

\bibitem[{{Abel} \emph{et~al.}(2002){Abel}, {Bryan} and {Norman}}]{Abel2002}
{Abel}, T., {Bryan}, G.~L.,  and {Norman}, M.~L. (2002). \enquote{{The
  Formation of the First Star in the Universe},} \emph{Science} \textbf{295},
  93--98.

\bibitem[{{Abel} \emph{et~al.}(2007){Abel}, {Wise} and {Bryan}}]{Abel2007}
{Abel}, T., {Wise}, J.~H.,  and {Bryan}, G.~L. (2007). \enquote{{The H II
  Region of a Primordial Star},} \emph{\apjl} \textbf{659},  L87--L90.

\bibitem[{{Agarwal} \emph{et~al.}(2012){Agarwal}, {Khochfar}, {Johnson},
  {Neistein}, {Dalla Vecchia} and {Livio}}]{Agarwal12}
{Agarwal}, B., {Khochfar}, S., {Johnson}, J.~L., {Neistein}, E., {Dalla
  Vecchia}, C.,  and {Livio}, M. (2012). \enquote{{Ubiquitous seeding of
  supermassive black holes by direct collapse},} \emph{\mnras} \textbf{425},
  2854--2871.

\bibitem[{{Agarwal} \emph{et~al.}(2016){Agarwal}, {Smith}, {Glover},
  {Natarajan} and {Khochfar}}]{Agarwal2016}
{Agarwal}, B., {Smith}, B., {Glover}, S., {Natarajan}, P.,  and {Khochfar}, S.
  (2016). \enquote{{New constraints on direct collapse black hole formation in
  the early Universe},} \emph{\mnras} \textbf{459},  4209--4217.

\bibitem[{{Alvarez} \emph{et~al.}(2006){Alvarez}, {Bromm} and
  {Shapiro}}]{Alvarez2006}
{Alvarez}, M.~A., {Bromm}, V.,  and {Shapiro}, P.~R. (2006). \enquote{{The H II
  Region of the First Star},} \emph{\apj} \textbf{639},  621--632.

\bibitem[{{Aoki} \emph{et~al.}(2014){Aoki}, {Tominaga}, {Beers}, {Honda} and
  {Lee}}]{Aoki2014}
{Aoki}, W., {Tominaga}, N., {Beers}, T.~C., {Honda}, S.,  and {Lee}, Y.~S.
  (2014). \enquote{{A chemical signature of first-generation very massive
  stars},} \emph{Science} \textbf{345},  912--915.

\bibitem[{{Asplund} \emph{et~al.}(2009){Asplund}, {Grevesse}, {Sauval} and
  {Scott}}]{Asplund2009}
{Asplund}, M., {Grevesse}, N., {Sauval}, A.~J.,  and {Scott}, P. (2009).
  \enquote{{The Chemical Composition of the Sun},} \emph{\araa} \textbf{47},
  481--522.

\bibitem[{{Balbus} and {Hawley}(1998)}]{Balbus1998}
{Balbus}, S.~A. and {Hawley}, J.~F. (1998). \enquote{{Instability, turbulence,
  and enhanced transport in accretion disks},} \emph{Reviews of Modern Physics}
  \textbf{70},  1--53.

\bibitem[{{Banerjee} and {Jedamzik}(2003)}]{Banerjee2003}
{Banerjee}, R. and {Jedamzik}, K. (2003). \enquote{{Are Cluster Magnetic Fields
  Primordial?}} \emph{Physical Review Letters} \textbf{91}, 25, 251301.

\bibitem[{{Barkana} and {Loeb}(2001)}]{Barkana2001}
{Barkana}, R. and {Loeb}, A. (2001). \enquote{{In the beginning: the first
  sources of light and the reionization of the universe},} \emph{\physrep}
  \textbf{349},  125--238.

\bibitem[{{Bate} \emph{et~al.}(1995){Bate}, {Bonnell} and {Price}}]{Bate1995}
{Bate}, M.~R., {Bonnell}, I.~A.,  and {Price}, N.~M. (1995).
  \enquote{{Modelling accretion in protobinary systems},} \emph{\mnras}
  \textbf{277},  362--376.

\bibitem[{{Beck} \emph{et~al.}(1996){Beck}, {Brandenburg}, {Moss}, {Shukurov}
  and {Sokoloff}}]{Beck1996}
{Beck}, R., {Brandenburg}, A., {Moss}, D., {Shukurov}, A.,  and {Sokoloff}, D.
  (1996). \enquote{{Galactic Magnetism: Recent Developments and Perspectives},}
  \emph{\araa} \textbf{34},  155--206.

\bibitem[{{Beers} and {Christlieb}(2005)}]{Beers2005}
{Beers}, T.~C. and {Christlieb}, N. (2005). \enquote{{The Discovery and
  Analysis of Very Metal-Poor Stars in the Galaxy},} \emph{\araa} \textbf{43},
  531--580.

\bibitem[{{Belczynski} \emph{et~al.}(2017){Belczynski}, {Ryu}, {Perna},
  {Berti}, {Tanaka} and {Bulik}}]{Belczynski2017}
{Belczynski}, K., {Ryu}, T., {Perna}, R., {Berti}, E., {Tanaka}, T.~L.,  and
  {Bulik}, T. (2017). \enquote{{On the likelihood of detecting gravitational
  waves from Population III compact object binaries},} \emph{\mnras}
  \textbf{471},  4702--4721.

\bibitem[{{Bertone} \emph{et~al.}(2005){Bertone}, {Hooper} and
  {Silk}}]{Bertone2005}
{Bertone}, G., {Hooper}, D.,  and {Silk}, J. (2005). \enquote{{Particle dark
  matter: evidence, candidates and constraints},} \emph{\physrep} \textbf{405},
   279--390.

\bibitem[{{Biermann}(1950)}]{Biermann1950}
{Biermann}, L. (1950). \enquote{{{\"U}ber den Ursprung der Magnetfelder auf
  Sternen und im interstellaren Raum (miteinem Anhang von A. Schl{\"u}ter)},}
  \emph{Zeitschrift Naturforschung Teil A} \textbf{5}, ~65.

\bibitem[{{Binney} and {Tremaine}(1987)}]{Binney1987}
{Binney}, J. and {Tremaine}, S. (1987). \emph{{Galactic dynamics}} (Princeton
  University Press).

\bibitem[{{Bleuler} and {Teyssier}(2014)}]{Bleuler2014}
{Bleuler}, A. and {Teyssier}, R. (2014). \enquote{{Towards a more realistic
  sink particle algorithm for the RAMSES CODE},} \emph{\mnras} \textbf{445},
  4015--4036.

\bibitem[{{Blumenthal} \emph{et~al.}(1986){Blumenthal}, {Faber}, {Flores} and
  {Primack}}]{Blumenthal1986}
{Blumenthal}, G.~R., {Faber}, S.~M., {Flores}, R.,  and {Primack}, J.~R.
  (1986). \enquote{{Contraction of dark matter galactic halos due to baryonic
  infall},} \emph{\apj} \textbf{301},  27--34.

\bibitem[{{Bonifacio} \emph{et~al.}(2015){Bonifacio}, {Caffau}, {Spite},
  {Limongi}, {Chieffi}, {Klessen}, {Fran{\c c}ois}, {Molaro}, {Ludwig},
  {Zaggia} and {Spite, F.~et~al.}}]{Bonifacio2015}
{Bonifacio}, P., {Caffau}, E., {Spite}, M., {Limongi}, M., {Chieffi}, A.,
  {Klessen}, R.~S., {Fran{\c c}ois}, P., {Molaro}, P., {Ludwig}, H.-G.,
  {Zaggia}, S.,  and {Spite, F.~et~al.} (2015). \enquote{{TOPoS . II. On the
  bimodality of carbon abundance in CEMP stars Implications on the early
  chemical evolution of galaxies},} \emph{\aap} \textbf{579}, A28.

\bibitem[{{Bovino} \emph{et~al.}(2016){Bovino}, {Grassi}, {Schleicher} and
  {Banerjee}}]{Bovino2016}
{Bovino}, S., {Grassi}, T., {Schleicher}, D.~R.~G.,  and {Banerjee}, R. (2016).
  \enquote{{The Formation of the Primitive Star SDSS J102915+172927: Effect of
  the Dust Mass and the Grain-size Distribution},} \emph{\apj} \textbf{832},
  154.

\bibitem[{{Bovino} \emph{et~al.}(2013){Bovino}, {Schleicher} and
  {Schober}}]{Bovino2013NJP}
{Bovino}, S., {Schleicher}, D.~R.~G.,  and {Schober}, J. (2013).
  \enquote{{Turbulent magnetic field amplification from the smallest to the
  largest magnetic Prandtl numbers},} \emph{New Journal of Physics}
  \textbf{15}, 1, 013055.

\bibitem[{{Brandenburg} and {Subramanian}(2005)}]{Brandenburg2005}
{Brandenburg}, A. and {Subramanian}, K. (2005). \enquote{{Astrophysical
  magnetic fields and nonlinear dynamo theory},} \emph{\physrep} \textbf{417},
  1--209.

\bibitem[{{Bromm}(2013)}]{Bromm2013}
{Bromm}, V. (2013). \enquote{{Formation of the first stars},} \emph{Reports on
  Progress in Physics} \textbf{76}, 11, 112901.

\bibitem[{{Bromm} \emph{et~al.}(2002){Bromm}, {Coppi} and {Larson}}]{Bromm2002}
{Bromm}, V., {Coppi}, P.~S.,  and {Larson}, R.~B. (2002). \enquote{{The
  Formation of the First Stars. I. The Primordial Star-forming Cloud},}
  \emph{\apj} \textbf{564},  23--51.

\bibitem[{{Bromm} and {Larson}(2004)}]{Bromm04}
{Bromm}, V. and {Larson}, R.~B. (2004). \enquote{{The First Stars},}
  \emph{\araa} \textbf{42},  79--118.

\bibitem[{{Bromm} and {Loeb}(2003)}]{Bromm03}
{Bromm}, V. and {Loeb}, A. (2003). \enquote{{Formation of the First
  Supermassive Black Holes},} \emph{\apj} \textbf{596},  34--46.

\bibitem[{{Bromm} and {Yoshida}(2011)}]{Bromm2011A}
{Bromm}, V. and {Yoshida}, N. (2011). \enquote{{The First Galaxies},}
  \emph{\araa} \textbf{49},  373--407.

\bibitem[{{Caffau} \emph{et~al.}(2011{\natexlab{a}}){Caffau}, {Bonifacio},
  {Fran{\c c}ois}, {Sbordone}, {Monaco}, {Spite}, {Spite}, {Ludwig}, {Cayrel},
  {Zaggia}, {Hammer}, {Randich}, {Molaro} and {Hill}}]{Caffau2011Nat}
{Caffau}, E., {Bonifacio}, P., {Fran{\c c}ois}, P., {Sbordone}, L., {Monaco},
  L., {Spite}, M., {Spite}, F., {Ludwig}, H.-G., {Cayrel}, R., {Zaggia}, S.,
  {Hammer}, F., {Randich}, S., {Molaro}, P.,  and {Hill}, V.
  (2011{\natexlab{a}}). \enquote{{An extremely primitive star in the Galactic
  halo},} \emph{\nat} \textbf{477},  67--69.

\bibitem[{{Caffau} \emph{et~al.}(2012){Caffau}, {Bonifacio}, {Fran{\c c}ois},
  {Spite}, {Spite}, {Zaggia}, {Ludwig}, {Steffen}, {Mashonkina}, {Monaco} and
  {Sbordone, L.~et~al.}}]{Caffau2012}
{Caffau}, E., {Bonifacio}, P., {Fran{\c c}ois}, P., {Spite}, M., {Spite}, F.,
  {Zaggia}, S., {Ludwig}, H.-G., {Steffen}, M., {Mashonkina}, L., {Monaco}, L.,
   and {Sbordone, L.~et~al.} (2012). \enquote{{A primordial star in the heart
  of the Lion},} \emph{\aap} \textbf{542}, A51.

\bibitem[{{Caffau} \emph{et~al.}(2013){Caffau}, {Bonifacio}, {Sbordone},
  {Fran{\c c}ois}, {Monaco}, {Spite}, {Plez}, {Cayrel}, {Christlieb}, {Clark},
  {Glover} and {Klessen, R.~et~al.}}]{Caffau2013}
{Caffau}, E., {Bonifacio}, P., {Sbordone}, L., {Fran{\c c}ois}, P., {Monaco},
  L., {Spite}, M., {Plez}, B., {Cayrel}, R., {Christlieb}, N., {Clark}, P.,
  {Glover}, S.,  and {Klessen, R.~et~al.} (2013). \enquote{{TOPoS. I. Survey
  design and analysis of the first sample},} \emph{\aap} \textbf{560}, A71.

\bibitem[{{Caffau} \emph{et~al.}(2011{\natexlab{b}}){Caffau}, {Ludwig},
  {Steffen}, {Freytag} and {Bonifacio}}]{Caffau2011}
{Caffau}, E., {Ludwig}, H.-G., {Steffen}, M., {Freytag}, B.,  and {Bonifacio},
  P. (2011{\natexlab{b}}). \enquote{{Solar Chemical Abundances Determined with
  a CO5BOLD 3D Model Atmosphere},} \emph{\solphys} \textbf{268},  255--269.

\bibitem[{{Chabrier}(2003)}]{Chabrier2003}
{Chabrier}, G. (2003). \enquote{{Galactic Stellar and Substellar Initial Mass
  Function},} \emph{\pasp} \textbf{115},  763--795.

\bibitem[{{Chan} and {Heger}(2017)}]{Chan2017}
{Chan}, C. and {Heger}, A. (2017). \enquote{{Combined Nucleosynthetic Yields of
  Multiple First Stars},} in S.~{Kubono}, T.~{Kajino}, S.~{Nishimura},
  T.~{Isobe}, S.~{Nagataki}, T.~{Shima},  and Y.~{Takeda} (eds.), \emph{14th
  International Symposium on Nuclei in the Cosmos (NIC2016)},  020209,
  \doi{10.7566/JPSCP.14.020209},
  \href{http://arxiv.org/abs/1610.06339}{\UrlFont{arXiv:1610.06339
  [astro-ph.SR]}}.

\bibitem[{{Chandrasekhar}(1951{\natexlab{a}})}]{Chandrasekhar1951A}
{Chandrasekhar}, S. (1951{\natexlab{a}}). \enquote{{The Fluctuations of Density
  in Isotropic Turbulence},} \emph{Proceedings of the Royal Society of London
  Series A} \textbf{210},  18--25.

\bibitem[{{Chandrasekhar}(1951{\natexlab{b}})}]{Chandrasekhar1951B}
{Chandrasekhar}, S. (1951{\natexlab{b}}). \enquote{{The Gravitational
  Instability of an Infinite Homogeneous Turbulent Medium},} \emph{Proceedings
  of the Royal Society of London Series A} \textbf{210},  26--29.

\bibitem[{{Chandrasekhar}(1964)}]{Chandra1964}
{Chandrasekhar}, S. (1964). \enquote{{The Dynamical Instability of Gaseous
  Masses Approaching the Schwarzschild Limit in General Relativity.}}
  \emph{\apj} \textbf{140},  417.

\bibitem[{{Chiaki} \emph{et~al.}(2013{\natexlab{a}}){Chiaki}, {Nozawa} and
  {Yoshida}}]{Chiaki2013B}
{Chiaki}, G., {Nozawa}, T.,  and {Yoshida}, N. (2013{\natexlab{a}}).
  \enquote{{Growth of Dust Grains in a Low-metallicity Gas and Its Effect on
  the Cloud Fragmentation},} \emph{\apjl} \textbf{765}, L3.

\bibitem[{{Chiaki} \emph{et~al.}(2014){Chiaki}, {Schneider}, {Nozawa},
  {Omukai}, {Limongi}, {Yoshida} and {Chieffi}}]{Chiaki2014}
{Chiaki}, G., {Schneider}, R., {Nozawa}, T., {Omukai}, K., {Limongi}, M.,
  {Yoshida}, N.,  and {Chieffi}, A. (2014). \enquote{{Dust grain growth and the
  formation of the extremely primitive star SDSS J102915+172927},}
  \emph{\mnras} \textbf{439},  3121--3127.

\bibitem[{{Chiaki} \emph{et~al.}(2016){Chiaki}, {Yoshida} and
  {Hirano}}]{Chiaki2016}
{Chiaki}, G., {Yoshida}, N.,  and {Hirano}, S. (2016). \enquote{{Gravitational
  collapse and the thermal evolution of low-metallicity gas clouds in the early
  Universe},} \emph{\mnras} \textbf{463},  2781--2798.

\bibitem[{{Chiaki} \emph{et~al.}(2013{\natexlab{b}}){Chiaki}, {Yoshida} and
  {Kitayama}}]{Chiaki2013}
{Chiaki}, G., {Yoshida}, N.,  and {Kitayama}, T. (2013{\natexlab{b}}).
  \enquote{{Low-mass Star Formation Triggered by Early Supernova Explosions},}
  \emph{\apj} \textbf{762}, 50.

\bibitem[{{Clark} \emph{et~al.}(2008){Clark}, {Glover} and
  {Klessen}}]{Clark2008}
{Clark}, P.~C., {Glover}, S.~C.~O.,  and {Klessen}, R.~S. (2008). \enquote{{The
  First Stellar Cluster},} \emph{\apj} \textbf{672}, 757-764.

\bibitem[{{Clark} \emph{et~al.}(2011{\natexlab{a}}){Clark}, {Glover}, {Klessen}
  and {Bromm}}]{Clark2011b}
{Clark}, P.~C., {Glover}, S.~C.~O., {Klessen}, R.~S.,  and {Bromm}, V.
  (2011{\natexlab{a}}). \enquote{{Gravitational Fragmentation in Turbulent
  Primordial Gas and the Initial Mass Function of Population III Stars},}
  \emph{\apj} \textbf{727}, 110.

\bibitem[{{Clark} \emph{et~al.}(2011{\natexlab{b}}){Clark}, {Glover}, {Smith},
  {Greif}, {Klessen} and {Bromm}}]{Clark11}
{Clark}, P.~C., {Glover}, S.~C.~O., {Smith}, R.~J., {Greif}, T.~H., {Klessen},
  R.~S.,  and {Bromm}, V. (2011{\natexlab{b}}). \enquote{{The Formation and
  Fragmentation of Disks Around Primordial Protostars},} \emph{Science}
  \textbf{331},  1040--.

\bibitem[{{Commer{\c c}on} \emph{et~al.}(2011){Commer{\c c}on}, {Hennebelle}
  and {Henning}}]{Commercon2011}
{Commer{\c c}on}, B., {Hennebelle}, P.,  and {Henning}, T. (2011).
  \enquote{{Collapse of Massive Magnetized Dense Cores Using Radiation
  Magnetohydrodynamics: Early Fragmentation Inhibition},} \emph{\apjl}
  \textbf{742}, L9.

\bibitem[{{Cooke} and {Madau}(2014)}]{Cooke2014}
{Cooke}, R.~J. and {Madau}, P. (2014). \enquote{{Carbon-enhanced Metal-poor
  Stars: Relics from the Dark Ages},} \emph{\apj} \textbf{791}, 116.

\bibitem[{{de Bennassuti} \emph{et~al.}(2014){de Bennassuti}, {Schneider},
  {Valiante} and {Salvadori}}]{Bennassuti2014}
{de Bennassuti}, M., {Schneider}, R., {Valiante}, R.,  and {Salvadori}, S.
  (2014). \enquote{{Decoding the stellar fossils of the dusty Milky Way
  progenitors},} \emph{\mnras} \textbf{445},  3039--3054.

\bibitem[{{Dopcke} \emph{et~al.}(2013){Dopcke}, {Glover}, {Clark} and
  {Klessen}}]{Dopcke2013}
{Dopcke}, G., {Glover}, S.~C.~O., {Clark}, P.~C.,  and {Klessen}, R.~S. (2013).
  \enquote{{On the Initial Mass Function of Low-metallicity Stars: The
  Importance of Dust Cooling},} \emph{\apj} \textbf{766}, 103.

\bibitem[{{Elmegreen}(2002)}]{Elmegreen2002}
{Elmegreen}, B.~G. (2002). \enquote{{Star Formation from Galaxies to
  Globules},} \emph{\apj} \textbf{577},  206--220.

\bibitem[{{Federrath} and {Klessen}(2012)}]{Federrath2012}
{Federrath}, C. and {Klessen}, R.~S. (2012). \enquote{{The Star Formation Rate
  of Turbulent Magnetized Clouds: Comparing Theory, Simulations, and
  Observations},} \emph{\apj} \textbf{761}, 156.

\bibitem[{{Federrath} \emph{et~al.}(2010){Federrath}, {Roman-Duval}, {Klessen},
  {Schmidt} and {Mac Low}}]{Federrath2010}
{Federrath}, C., {Roman-Duval}, J., {Klessen}, R.~S., {Schmidt}, W.,  and {Mac
  Low}, M.-M. (2010). \enquote{{Comparing the statistics of interstellar
  turbulence in simulations and observations. Solenoidal versus compressive
  turbulence forcing},} \emph{\aap} \textbf{512}, A81.

\bibitem[{{Federrath} \emph{et~al.}(2014){Federrath}, {Schober}, {Bovino} and
  {Schleicher}}]{Federrath2014}
{Federrath}, C., {Schober}, J., {Bovino}, S.,  and {Schleicher}, D.~R.~G.
  (2014). \enquote{{The Turbulent Dynamo in Highly Compressible Supersonic
  Plasmas},} \emph{\apjl} \textbf{797}, L19.

\bibitem[{{Federrath} \emph{et~al.}(2011){Federrath}, {Sur}, {Schleicher},
  {Banerjee} and {Klessen}}]{Federrath2011}
{Federrath}, C., {Sur}, S., {Schleicher}, D.~R.~G., {Banerjee}, R.,  and
  {Klessen}, R.~S. (2011). \enquote{{A New Jeans Resolution Criterion for (M)HD
  Simulations of Self-gravitating Gas: Application to Magnetic Field
  Amplification by Gravity-driven Turbulence},} \emph{\apj} \textbf{731}, 62.

\bibitem[{{Ferri{\`e}re} \emph{et~al.}(2007){Ferri{\`e}re}, {Gillard} and
  {Jean}}]{Feri2007}
{Ferri{\`e}re}, K., {Gillard}, W.,  and {Jean}, P. (2007). \enquote{{Spatial
  distribution of interstellar gas in the innermost 3 kpc of our galaxy},}
  \emph{\aap} \textbf{467},  611--627.

\bibitem[{{Fialkov} \emph{et~al.}(2012){Fialkov}, {Barkana}, {Tseliakhovich}
  and {Hirata}}]{Fialkov2012}
{Fialkov}, A., {Barkana}, R., {Tseliakhovich}, D.,  and {Hirata}, C.~M. (2012).
  \enquote{{Impact of the relative motion between the dark matter and baryons
  on the first stars: semi-analytical modelling},} \emph{\mnras} \textbf{424},
  1335--1345.

\bibitem[{{Fialkov} \emph{et~al.}(2014){Fialkov}, {Barkana} and
  {Visbal}}]{Fialkov14}
{Fialkov}, A., {Barkana}, R.,  and {Visbal}, E. (2014). \enquote{{The
  observable signature of late heating of the Universe during cosmic
  reionization},} \emph{\nat} \textbf{506},  197--199.

\bibitem[{{Frebel}(2010)}]{Frebel2010}
{Frebel}, A. (2010). \enquote{{Stellar archaeology: Exploring the Universe with
  metal-poor stars},} \emph{Astronomische Nachrichten} \textbf{331},  474--488.

\bibitem[{{Frebel} \emph{et~al.}(2005){Frebel}, {Aoki}, {Christlieb}, {Ando},
  {Asplund}, {Barklem}, {Beers}, {Eriksson}, {Fechner}, {Fujimoto}, {Honda} and
  {Kajino, T.~et~al.}}]{Frebel2005}
{Frebel}, A., {Aoki}, W., {Christlieb}, N., {Ando}, H., {Asplund}, M.,
  {Barklem}, P.~S., {Beers}, T.~C., {Eriksson}, K., {Fechner}, C., {Fujimoto},
  M.~Y., {Honda}, S.,  and {Kajino, T.~et~al.} (2005).
  \enquote{{Nucleosynthetic signatures of the first stars},} \emph{\nat}
  \textbf{434},  871--873.

\bibitem[{{Frebel} \emph{et~al.}(2009){Frebel}, {Johnson} and
  {Bromm}}]{Frebel2009}
{Frebel}, A., {Johnson}, J.~L.,  and {Bromm}, V. (2009). \enquote{{The minimum
  stellar metallicity observable in the Galaxy},} \emph{\mnras} \textbf{392},
  L50--L54.

\bibitem[{{Freese} \emph{et~al.}(2009){Freese}, {Bodenheimer}, {Gondolo} and
  {Spolyar}}]{Freese2009}
{Freese}, K., {Bodenheimer}, P., {Gondolo}, P.,  and {Spolyar}, D. (2009).
  \enquote{{Dark stars: a new study of the first stars in the Universe},}
  \emph{New Journal of Physics} \textbf{11}, 10, 105014.

\bibitem[{{Freese} \emph{et~al.}(2008){Freese}, {Bodenheimer}, {Spolyar} and
  {Gondolo}}]{Freese2008}
{Freese}, K., {Bodenheimer}, P., {Spolyar}, D.,  and {Gondolo}, P. (2008).
  \enquote{{Stellar Structure of Dark Stars: A First Phase of Stellar Evolution
  Resulting from Dark Matter Annihilation},} \emph{\apjl} \textbf{685}, L101.

\bibitem[{{Gammie}(2001)}]{Gammie01}
{Gammie}, C.~F. (2001). \enquote{{Nonlinear Outcome of Gravitational
  Instability in Cooling, Gaseous Disks},} \emph{\apj} \textbf{553},  174--183.

\bibitem[{Gardner \emph{et~al.}(2006)Gardner, Mather, Clampin, Doyon,
  Greenhouse, Hammel, Hutchings, Jakobsen, Lilly and Long}]{Gardner2006}
Gardner, J.~P., Mather, J.~C., Clampin, M., Doyon, R., Greenhouse, M.~A.,
  Hammel, H.~B., Hutchings, J.~B., Jakobsen, P., Lilly, S.~J.,  and Long, K.
  S.~e.~a. (2006). \enquote{{The James Webb Space Telescope},} \emph{Space
  Science Reviews} .

\bibitem[{{Girichidis} \emph{et~al.}(2012{\natexlab{a}}){Girichidis},
  {Federrath}, {Allison}, {Banerjee} and {Klessen}}]{Girichidis2012B}
{Girichidis}, P., {Federrath}, C., {Allison}, R., {Banerjee}, R.,  and
  {Klessen}, R.~S. (2012{\natexlab{a}}). \enquote{{Importance of the initial
  conditions for star formation - III. Statistical properties of embedded
  protostellar clusters},} \emph{\mnras} \textbf{420},  3264--3280.

\bibitem[{{Girichidis} \emph{et~al.}(2011){Girichidis}, {Federrath}, {Banerjee}
  and {Klessen}}]{Girichidis2011}
{Girichidis}, P., {Federrath}, C., {Banerjee}, R.,  and {Klessen}, R.~S.
  (2011). \enquote{{Importance of the initial conditions for star formation -
  I. Cloud evolution and morphology},} \emph{\mnras} \textbf{413},  2741--2759.

\bibitem[{{Girichidis} \emph{et~al.}(2012{\natexlab{b}}){Girichidis},
  {Federrath}, {Banerjee} and {Klessen}}]{Girichidis2012}
{Girichidis}, P., {Federrath}, C., {Banerjee}, R.,  and {Klessen}, R.~S.
  (2012{\natexlab{b}}). \enquote{{Importance of the initial conditions for star
  formation - II. Fragmentation-induced starvation and accretion shielding},}
  \emph{\mnras} \textbf{420},  613--626.

\bibitem[{{Glover}(2005)}]{Glover2005}
{Glover}, S. (2005). \enquote{{The Formation Of The First Stars In The
  Universe},} \emph{\ssr} \textbf{117},  445--508.

\bibitem[{{Glover}(2013)}]{Glover2013}
{Glover}, S. (2013). \enquote{{The First Stars},} in T.~{Wiklind},
  B.~{Mobasher},  and V.~{Bromm} (eds.), \emph{The First Galaxies},
  \emph{Astrophysics and Space Science Library}, Vol. 396,  103,
  \doi{10.1007/978-3-642-32362-1_3},
  \href{http://arxiv.org/abs/1209.2509}{\UrlFont{arXiv:1209.2509}}.

\bibitem[{{Glover} and {Clark}(2014)}]{Glover2014}
{Glover}, S.~C.~O. and {Clark}, P.~C. (2014). \enquote{{Molecular cooling in
  the diffuse interstellar medium},} \emph{\mnras} \textbf{437},  9--20.

\bibitem[{{Grasso} and {Rubinstein}(2001)}]{Grasso2001}
{Grasso}, D. and {Rubinstein}, H.~R. (2001). \enquote{{Magnetic fields in the
  early Universe},} \emph{\physrep} \textbf{348},  163--266.

\bibitem[{{Greif}(2014)}]{Greif2014}
{Greif}, T.~H. (2014). \enquote{{The numerical frontier of the high-redshift
  Universe},} \emph{ArXiv e-prints:1410.3482} .

\bibitem[{{Greif}(2015)}]{Greif15}
{Greif}, T.~H. (2015). \enquote{{The numerical frontier of the high-redshift
  Universe},} \emph{Computational Astrophysics and Cosmology} \textbf{2}, 3.

\bibitem[{{Greif} \emph{et~al.}(2012){Greif}, {Bromm}, {Clark}, {Glover},
  {Smith}, {Klessen}, {Yoshida} and {Springel}}]{Greif12}
{Greif}, T.~H., {Bromm}, V., {Clark}, P.~C., {Glover}, S.~C.~O., {Smith},
  R.~J., {Klessen}, R.~S., {Yoshida}, N.,  and {Springel}, V. (2012).
  \enquote{{Formation and evolution of primordial protostellar systems},}
  \emph{\mnras} \textbf{424},  399--415.

\bibitem[{{Greif} \emph{et~al.}(2008){Greif}, {Johnson}, {Klessen} and
  {Bromm}}]{Greif2008}
{Greif}, T.~H., {Johnson}, J.~L., {Klessen}, R.~S.,  and {Bromm}, V. (2008).
  \enquote{{The first galaxies: assembly, cooling and the onset of
  turbulence},} \emph{\mnras} \textbf{387},  1021--1036.

\bibitem[{{Greif} \emph{et~al.}(2011){Greif}, {White}, {Klessen} and
  {Springel}}]{Greif2011}
{Greif}, T.~H., {White}, S.~D.~M., {Klessen}, R.~S.,  and {Springel}, V.
  (2011). \enquote{{The Delay of Population III Star Formation by Supersonic
  Streaming Velocities},} \emph{\apj} \textbf{736}, 147.

\bibitem[{{Griffen} \emph{et~al.}(2018){Griffen}, {Dooley}, {Ji}, {O'Shea},
  {G{\'o}mez} and {Frebel}}]{Griffen16}
{Griffen}, B.~F., {Dooley}, G.~A., {Ji}, A.~P., {O'Shea}, B.~W., {G{\'o}mez},
  F.~A.,  and {Frebel}, A. (2018). \enquote{{Tracing the first stars and
  galaxies of the Milky Way},} \emph{\mnras} \textbf{474},  443--459.

\bibitem[{{Haemmerl{\'e}} \emph{et~al.}(2016){Haemmerl{\'e}}, {Eggenberger},
  {Meynet}, {Maeder} and {Charbonnel}}]{Haemmerle16}
{Haemmerl{\'e}}, L., {Eggenberger}, P., {Meynet}, G., {Maeder}, A.,  and
  {Charbonnel}, C. (2016). \enquote{{Massive star formation by accretion. I.
  Disc accretion},} \emph{A\&A} \textbf{585}, A65.

\bibitem[{{Hartwig} \emph{et~al.}(2015){Hartwig}, {Bromm}, {Klessen} and
  {Glover}}]{Hartwig2015MNRAS}
{Hartwig}, T., {Bromm}, V., {Klessen}, R.~S.,  and {Glover}, S.~C.~O. (2015).
  \enquote{{Constraining the primordial initial mass function with stellar
  archaeology},} \emph{\mnras} \textbf{447},  3892--3908.

\bibitem[{{Hartwig} \emph{et~al.}(2016){Hartwig}, {Volonteri}, {Bromm},
  {Klessen}, {Barausse}, {Magg} and {Stacy}}]{Hartwig2016b}
{Hartwig}, T., {Volonteri}, M., {Bromm}, V., {Klessen}, R.~S., {Barausse}, E.,
  {Magg}, M.,  and {Stacy}, A. (2016). \enquote{{Gravitational waves from the
  remnants of the first stars},} \emph{\mnras} \textbf{460},  L74--L78.

\bibitem[{{Heger} and {Woosley}(2002)}]{Heger2002}
{Heger}, A. and {Woosley}, S.~E. (2002). \enquote{{The Nucleosynthetic
  Signature of Population III},} \emph{\apj} \textbf{567},  532--543.

\bibitem[{{Heger} and {Woosley}(2010)}]{Heger2010}
{Heger}, A. and {Woosley}, S.~E. (2010). \enquote{{Nucleosynthesis and
  Evolution of Massive Metal-free Stars},} \emph{\apj} \textbf{724},  341--373.

\bibitem[{{Hirano} \emph{et~al.}(2015){Hirano}, {Hosokawa}, {Yoshida}, {Omukai}
  and {Yorke}}]{Hirano2015}
{Hirano}, S., {Hosokawa}, T., {Yoshida}, N., {Omukai}, K.,  and {Yorke}, H.~W.
  (2015). \enquote{{Primordial star formation under the influence of far
  ultraviolet radiation: 1540 cosmological haloes and the stellar mass
  distribution},} \emph{\mnras} \textbf{448},  568--587.

\bibitem[{{Hirano} \emph{et~al.}(2014){Hirano}, {Hosokawa}, {Yoshida}, {Umeda},
  {Omukai}, {Chiaki} and {Yorke}}]{Hirano2014}
{Hirano}, S., {Hosokawa}, T., {Yoshida}, N., {Umeda}, H., {Omukai}, K.,
  {Chiaki}, G.,  and {Yorke}, H.~W. (2014). \enquote{{One Hundred First Stars:
  Protostellar Evolution and the Final Masses},} \emph{\apj} \textbf{781}, 60.

\bibitem[{{Hirano} \emph{et~al.}(2011){Hirano}, {Umeda} and
  {Yoshida}}]{Hirano2011}
{Hirano}, S., {Umeda}, H.,  and {Yoshida}, N. (2011). \enquote{{Evolution of
  Primordial Stars Powered by Dark Matter Annihilation up to the Main-sequence
  Stage},} \emph{\apj} \textbf{736}, 58.

\bibitem[{{Hosokawa} \emph{et~al.}(2016){Hosokawa}, {Hirano}, {Kuiper},
  {Yorke}, {Omukai} and {Yoshida}}]{Hosokawa2016}
{Hosokawa}, T., {Hirano}, S., {Kuiper}, R., {Yorke}, H.~W., {Omukai}, K.,  and
  {Yoshida}, N. (2016). \enquote{{Formation of Massive Primordial Stars:
  Intermittent UV Feedback with Episodic Mass Accretion},} \emph{\apj}
  \textbf{824}, 119.

\bibitem[{{Hosokawa} and {Omukai}(2009)}]{Hosokawa2009}
{Hosokawa}, T. and {Omukai}, K. (2009). \enquote{{Evolution of Massive
  Protostars with High Accretion Rates},} \emph{\apj} \textbf{691},  823--846.

\bibitem[{{Hosokawa} \emph{et~al.}(2011){Hosokawa}, {Omukai}, {Yoshida} and
  {Yorke}}]{Hosokawa11}
{Hosokawa}, T., {Omukai}, K., {Yoshida}, N.,  and {Yorke}, H.~W. (2011).
  \enquote{{Protostellar Feedback Halts the Growth of the First Stars in the
  Universe},} \emph{Science} \textbf{334},  1250--.

\bibitem[{{Hosokawa} \emph{et~al.}(2013){Hosokawa}, {Yorke}, {Inayoshi},
  {Omukai} and {Yoshida}}]{Hosokawa2013}
{Hosokawa}, T., {Yorke}, H.~W., {Inayoshi}, K., {Omukai}, K.,  and {Yoshida},
  N. (2013). \enquote{{Formation of Primordial Supermassive Stars by Rapid Mass
  Accretion},} \emph{\apj} \textbf{778}, 178.

\bibitem[{{Hosokawa} \emph{et~al.}(2010){Hosokawa}, {Yorke} and
  {Omukai}}]{Hosokawa2010}
{Hosokawa}, T., {Yorke}, H.~W.,  and {Omukai}, K. (2010). \enquote{{Evolution
  of Massive Protostars Via Disk Accretion},} \emph{\apj} \textbf{721},
  478--492.

\bibitem[{{Hosokawa} \emph{et~al.}(2012){Hosokawa}, {Yoshida}, {Omukai} and
  {Yorke}}]{Hosokawa2012}
{Hosokawa}, T., {Yoshida}, N., {Omukai}, K.,  and {Yorke}, H.~W. (2012).
  \enquote{{Protostellar Feedback and Final Mass of the Second-generation
  Primordial Stars},} \emph{\apjl} \textbf{760}, L37.

\bibitem[{{Howard} \emph{et~al.}(2014){Howard}, {Pudritz} and
  {Harris}}]{Howard2014}
{Howard}, C.~S., {Pudritz}, R.~E.,  and {Harris}, W.~E. (2014).
  \enquote{{Cluster formation in molecular clouds - I. Stellar populations,
  star formation rates and ionizing radiation},} \emph{\mnras} \textbf{438},
  1305--1317.

\bibitem[{{Hummel} \emph{et~al.}(2012){Hummel}, {Pawlik}, {Milosavljevi{\'c}}
  and {Bromm}}]{Hummel2012}
{Hummel}, J.~A., {Pawlik}, A.~H., {Milosavljevi{\'c}}, M.,  and {Bromm}, V.
  (2012). \enquote{{The Source Density and Observability of Pair-instability
  Supernovae from the First Stars},} \emph{\apj} \textbf{755}, 72.

\bibitem[{{Iben}(1963)}]{Iben1963}
{Iben}, I., Jr. (1963). \enquote{{Massive Stars in Quasi-Static Equlibrium.}}
  \emph{\apj} \textbf{138},  1090.

\bibitem[{{Inayoshi} \emph{et~al.}(2016){Inayoshi}, {Haiman} and
  {Ostriker}}]{Kohei}
{Inayoshi}, K., {Haiman}, Z.,  and {Ostriker}, J.~P. (2016).
  \enquote{{Hyper-Eddington accretion flows on to massive black holes},}
  \emph{\mnras} \textbf{459},  3738--3755.

\bibitem[{{Iocco}(2008)}]{Iocco2008A}
{Iocco}, F. (2008). \enquote{{Dark Matter Capture and Annihilation on the First
  Stars: Preliminary Estimates},} \emph{\apjl} \textbf{677}, L1.

\bibitem[{{Iocco} \emph{et~al.}(2008){Iocco}, {Bressan}, {Ripamonti},
  {Schneider}, {Ferrara} and {Marigo}}]{Iocco2008B}
{Iocco}, F., {Bressan}, A., {Ripamonti}, E., {Schneider}, R., {Ferrara}, A.,
  and {Marigo}, P. (2008). \enquote{{Dark matter annihilation effects on the
  first stars},} \emph{\mnras} \textbf{390},  1655--1669.

\bibitem[{{Iwamoto} \emph{et~al.}(2005){Iwamoto}, {Umeda}, {Tominaga}, {Nomoto}
  and {Maeda}}]{Iwamoto2005}
{Iwamoto}, N., {Umeda}, H., {Tominaga}, N., {Nomoto}, K.,  and {Maeda}, K.
  (2005). \enquote{{The First Chemical Enrichment in the Universe and the
  Formation of Hyper Metal-Poor Stars},} \emph{Science} \textbf{309},
  451--453.

\bibitem[{{Jappsen} \emph{et~al.}(2009){Jappsen}, {Klessen}, {Glover} and {Mac
  Low}}]{Jappsen2009}
{Jappsen}, A., {Klessen}, R.~S., {Glover}, S.~C.~O.,  and {Mac Low}, M. (2009).
  \enquote{{Star Formation at Very Low Metallicity. IV. Fragmentation does not
  Depend on Metallicity for Cold Initial Conditions},} \emph{\apj}
  \textbf{696},  1065--1074.

\bibitem[{{Jappsen} \emph{et~al.}(2007){Jappsen}, {Glover}, {Klessen} and {Mac
  Low}}]{Jappsen2007}
{Jappsen}, A.-K., {Glover}, S.~C.~O., {Klessen}, R.~S.,  and {Mac Low}, M.-M.
  (2007). \enquote{{Star Formation at Very Low Metallicity. II. On the
  Insignificance of Metal-Line Cooling During the Early Stages of Gravitational
  Collapse},} \emph{\apj} \textbf{660},  1332--1343.

\bibitem[{{Jappsen} \emph{et~al.}(2005){Jappsen}, {Klessen}, {Larson}, {Li} and
  {Mac Low}}]{Jappsen2005}
{Jappsen}, A.-K., {Klessen}, R.~S., {Larson}, R.~B., {Li}, Y.,  and {Mac Low},
  M.-M. (2005). \enquote{{The stellar mass spectrum from non-isothermal
  gravoturbulent fragmentation},} \emph{\aap} \textbf{435},  611--623.

\bibitem[{{Jeans}(1902)}]{Jeans1902}
{Jeans}, J.~H. (1902). \enquote{{The Stability of a Spherical Nebula},}
  \emph{Philosophical Transactions of the Royal Society of London Series A}
  \textbf{199},  1--53.

\bibitem[{{Jeon} \emph{et~al.}(2014){Jeon}, {Pawlik}, {Bromm} and
  {Milosavljevi{\'c}}}]{Jeon2014}
{Jeon}, M., {Pawlik}, A.~H., {Bromm}, V.,  and {Milosavljevi{\'c}}, M. (2014).
  \enquote{{Radiative feedback from high-mass X-ray binaries on the formation
  of the first galaxies and early reionization},} \emph{\mnras} \textbf{440},
  3778--3796.

\bibitem[{{Ji} \emph{et~al.}(2016{\natexlab{a}}){Ji}, {Frebel}, {Ezzeddine} and
  {Casey}}]{Ji2016A}
{Ji}, A.~P., {Frebel}, A., {Ezzeddine}, R.,  and {Casey}, A.~R.
  (2016{\natexlab{a}}). \enquote{{Chemical Diversity in the Ultra-faint Dwarf
  Galaxy Tucana II},} \emph{\apjl} \textbf{832}, L3.

\bibitem[{{Ji} \emph{et~al.}(2016{\natexlab{b}}){Ji}, {Frebel}, {Simon} and
  {Chiti}}]{Ji2016B}
{Ji}, A.~P., {Frebel}, A., {Simon}, J.~D.,  and {Chiti}, A.
  (2016{\natexlab{b}}). \enquote{{Complete Element Abundances of Nine Stars in
  the r-process Galaxy Reticulum II},} \emph{\apj} \textbf{830}, 93.

\bibitem[{{Joggerst} \emph{et~al.}(2010){Joggerst}, {Almgren}, {Bell}, {Heger},
  {Whalen} and {Woosley}}]{Joggerst2010}
{Joggerst}, C.~C., {Almgren}, A., {Bell}, J., {Heger}, A., {Whalen}, D.,  and
  {Woosley}, S.~E. (2010). \enquote{{The Nucleosynthetic Imprint of 15-40 M
  $_{sun}$ Primordial Supernovae on Metal-Poor Stars},} \emph{\apj}
  \textbf{709},  11--26.

\bibitem[{{Joggerst} \emph{et~al.}(2009){Joggerst}, {Woosley} and
  {Heger}}]{Joggerst2009}
{Joggerst}, C.~C., {Woosley}, S.~E.,  and {Heger}, A. (2009). \enquote{{Mixing
  in Zero- and Solar-Metallicity Supernovae},} \emph{\apj} \textbf{693},
  1780--1802.

\bibitem[{{Kasen} \emph{et~al.}(2011){Kasen}, {Woosley} and
  {Heger}}]{Kasen2011}
{Kasen}, D., {Woosley}, S.~E.,  and {Heger}, A. (2011). \enquote{{Pair
  Instability Supernovae: Light Curves, Spectra, and Shock Breakout},}
  \emph{\apj} \textbf{734}, 102.

\bibitem[{{Kazantsev}(1968)}]{Kazantsev1968}
{Kazantsev}, A.~P. (1968). \enquote{{Enhancement of a Magnetic Field by a
  Conducting Fluid},} \emph{Soviet Journal of Experimental and Theoretical
  Physics} \textbf{26},  1031.

\bibitem[{{Keller} \emph{et~al.}(2014){Keller}, {Bessell}, {Frebel}, {Casey},
  {Asplund}, {Jacobson}, {Lind}, {Norris}, {Yong}, {Heger}, {Magic} and {da
  Costa, G.~S.~et~al.}}]{Keller2014}
{Keller}, S.~C., {Bessell}, M.~S., {Frebel}, A., {Casey}, A.~R., {Asplund}, M.,
  {Jacobson}, H.~R., {Lind}, K., {Norris}, J.~E., {Yong}, D., {Heger}, A.,
  {Magic}, Z.,  and {da Costa, G.~S.~et~al.} (2014). \enquote{{A single
  low-energy, iron-poor supernova as the source of metals in the star SMSS
  J031300.36-670839.3},} \emph{\nat} \textbf{506},  463--466.

\bibitem[{{Keto}(2007)}]{keto2007}
{Keto}, E. (2007). \enquote{{The Formation of Massive Stars: Accretion, Disks,
  and the Development of Hypercompact H II Regions},} \emph{\apj} \textbf{666},
   976--981.

\bibitem[{{Kippenhahn} \emph{et~al.}(2012){Kippenhahn}, {Weigert} and
  {Weiss}}]{Kippenhahn2012}
{Kippenhahn}, R., {Weigert}, A.,  and {Weiss}, A. (2012). \emph{{Stellar
  Structure and Evolution}} (Springer Verlag Berlin Heidelberg),
  \doi{10.1007/978-3-642-30304-3}.

\bibitem[{{Kirby} \emph{et~al.}(2015){Kirby}, {Guo}, {Zhang}, {Deng}, {Cohen},
  {Guhathakurta}, {Shetrone}, {Lee} and {Rizzi}}]{Kirby2015}
{Kirby}, E.~N., {Guo}, M., {Zhang}, A.~J., {Deng}, M., {Cohen}, J.~G.,
  {Guhathakurta}, P., {Shetrone}, M.~D., {Lee}, Y.~S.,  and {Rizzi}, L. (2015).
  \enquote{{Carbon in Red Giants in Globular Clusters and Dwarf Spheroidal
  Galaxies},} \emph{\apj} \textbf{801}, 125.

\bibitem[{{Kitayama} \emph{et~al.}(2004){Kitayama}, {Yoshida}, {Susa} and
  {Umemura}}]{Kitayama2004}
{Kitayama}, T., {Yoshida}, N., {Susa}, H.,  and {Umemura}, M. (2004).
  \enquote{{The Structure and Evolution of Early Cosmological H II Regions},}
  \emph{\apj} \textbf{613},  631--645.

\bibitem[{{Klessen} and {Glover}(2016)}]{Klessen2016}
{Klessen}, R.~S. and {Glover}, S.~C.~O. (2016). \enquote{{Physical Processes in
  the Interstellar Medium},} \emph{Star Formation in Galaxy Evolution:
  Connecting Numerical Models to Reality, Saas-Fee Advanced Course, Volume
  43.~ISBN 978-3-662-47889-9.~Springer-Verlag Berlin Heidelberg, 2016, p.~85}
  \textbf{43}, ~85.

\bibitem[{{Klessen} \emph{et~al.}(2012){Klessen}, {Glover} and
  {Clark}}]{Klessen2012}
{Klessen}, R.~S., {Glover}, S.~C.~O.,  and {Clark}, P.~C. (2012). \enquote{{On
  the formation of very metal poor stars: the case of SDSS J1029151+172927},}
  \emph{\mnras} \textbf{421},  3217--3221.

\bibitem[{{Koch} \emph{et~al.}(2013){Koch}, {Feltzing}, {Ad{\'e}n} and
  {Matteucci}}]{Koch2013}
{Koch}, A., {Feltzing}, S., {Ad{\'e}n}, D.,  and {Matteucci}, F. (2013).
  \enquote{{Neutron-capture element deficiency of the Hercules dwarf spheroidal
  galaxy},} \emph{\aap} \textbf{554}, A5.

\bibitem[{{Kratter} and {Lodato}(2016)}]{Kratter2016}
{Kratter}, K. and {Lodato}, G. (2016). \enquote{{Gravitational Instabilities in
  Circumstellar Disks},} \emph{\araa} \textbf{54},  271--311.

\bibitem[{{Kratter} and {Matzner}(2006)}]{Kratter2006}
{Kratter}, K.~M. and {Matzner}, C.~D. (2006). \enquote{{Fragmentation of
  massive protostellar discs},} \emph{\mnras} \textbf{373},  1563--1576.

\bibitem[{{Kratter} \emph{et~al.}(2010){Kratter}, {Murray-Clay} and
  {Youdin}}]{Kratter2010}
{Kratter}, K.~M., {Murray-Clay}, R.~A.,  and {Youdin}, A.~N. (2010).
  \enquote{{The Runts of the Litter: Why Planets Formed Through Gravitational
  Instability Can Only Be Failed Binary Stars},} \emph{\apj} \textbf{710},
  1375--1386.

\bibitem[{{Kroupa}(2002)}]{Kroupa2002}
{Kroupa}, P. (2002). \enquote{{The Initial Mass Function of Stars: Evidence for
  Uniformity in Variable Systems},} \emph{Science} \textbf{295},  82--91.

\bibitem[{{Krumholz}(2015)}]{Krumholz2015}
{Krumholz}, M.~R. (2015). \enquote{{Notes on Star Formation},} \emph{ArXiv
  e-prints:1511.03457} .

\bibitem[{{Krumholz} \emph{et~al.}(2009){Krumholz}, {Klein}, {McKee}, {Offner}
  and {Cunningham}}]{Krumholz2009}
{Krumholz}, M.~R., {Klein}, R.~I., {McKee}, C.~F., {Offner}, S.~S.~R.,  and
  {Cunningham}, A.~J. (2009). \enquote{{The Formation of Massive Star Systems
  by Accretion},} \emph{Science} \textbf{323},  754.

\bibitem[{{Krumholz} \emph{et~al.}(2004){Krumholz}, {McKee} and
  {Klein}}]{Krumholz2004}
{Krumholz}, M.~R., {McKee}, C.~F.,  and {Klein}, R.~I. (2004).
  \enquote{{Embedding Lagrangian Sink Particles in Eulerian Grids},}
  \emph{\apj} \textbf{611},  399--412.

\bibitem[{{Kuiper} \emph{et~al.}(2010){Kuiper}, {Klahr}, {Beuther} and
  {Henning}}]{Kuiper2010}
{Kuiper}, R., {Klahr}, H., {Beuther}, H.,  and {Henning}, T. (2010).
  \enquote{{Circumventing the Radiation Pressure Barrier in the Formation of
  Massive Stars via Disk Accretion},} \emph{\apj} \textbf{722},  1556--1576.

\bibitem[{{Kuiper} \emph{et~al.}(2011){Kuiper}, {Klahr}, {Beuther} and
  {Henning}}]{Kuiper2011}
{Kuiper}, R., {Klahr}, H., {Beuther}, H.,  and {Henning}, T. (2011).
  \enquote{{Three-dimensional Simulation of Massive Star Formation in the Disk
  Accretion Scenario},} \emph{\apj} \textbf{732}, 20.

\bibitem[{{Kulkarni} \emph{et~al.}(2014){Kulkarni}, {Hennawi}, {Rollinde} and
  {Vangioni}}]{Kulkarni2014}
{Kulkarni}, G., {Hennawi}, J.~F., {Rollinde}, E.,  and {Vangioni}, E. (2014).
  \enquote{{Chemical Constraints on the Contribution of Population III Stars to
  Cosmic Reionization},} \emph{\apj} \textbf{787}, 64.

\bibitem[{{Kulkarni} \emph{et~al.}(2013){Kulkarni}, {Rollinde}, {Hennawi} and
  {Vangioni}}]{Kulkarni2013}
{Kulkarni}, G., {Rollinde}, E., {Hennawi}, J.~F.,  and {Vangioni}, E. (2013).
  \enquote{{Chemical Enrichment of Damped Ly{$\alpha$} Systems as a Direct
  Constraint on Population III Star Formation},} \emph{\apj} \textbf{772}, 93.

\bibitem[{{Lada} and {Lada}(2003)}]{Lada2003}
{Lada}, C.~J. and {Lada}, E.~A. (2003). \enquote{{Embedded Clusters in
  Molecular Clouds},} \emph{\araa} \textbf{41},  57--115.

\bibitem[{{Lai} \emph{et~al.}(2008){Lai}, {Bolte}, {Johnson}, {Lucatello},
  {Heger} and {Woosley}}]{Lai2008}
{Lai}, D.~K., {Bolte}, M., {Johnson}, J.~A., {Lucatello}, S., {Heger}, A.,  and
  {Woosley}, S.~E. (2008). \enquote{{Detailed Abundances for 28 Metal-poor
  Stars: Stellar Relics in the Milky Way},} \emph{\apj} \textbf{681},
  1524-1556.

\bibitem[{{Larson}(1969)}]{Larson1969}
{Larson}, R.~B. (1969). \enquote{{Numerical calculations of the dynamics of
  collapsing proto-star},} \emph{\mnras} \textbf{145},  271.

\bibitem[{{Latif} \emph{et~al.}(2014{\natexlab{a}}){Latif}, {Niemeyer} and
  {Schleicher}}]{Latif2014Stream}
{Latif}, M.~A., {Niemeyer}, J.~C.,  and {Schleicher}, D.~R.~G.
  (2014{\natexlab{a}}). \enquote{{Impact of baryonic streaming velocities on
  the formation of supermassive black holes via direct collapse},}
  \emph{\mnras} \textbf{440},  2969--2975.

\bibitem[{{Latif} \emph{et~al.}(2014{\natexlab{b}}){Latif}, {Schleicher} and
  {Schmidt}}]{LatifMag2014}
{Latif}, M.~A., {Schleicher}, D.~R.~G.,  and {Schmidt}, W.
  (2014{\natexlab{b}}). \enquote{{Magnetic fields during the formation of
  supermassive black holes},} \emph{\mnras} \textbf{440},  1551--1561.

\bibitem[{{Latif} \emph{et~al.}(2013){Latif}, {Schleicher}, {Schmidt} and
  {Niemeyer}}]{Latif2013a}
{Latif}, M.~A., {Schleicher}, D.~R.~G., {Schmidt}, W.,  and {Niemeyer}, J.
  (2013). \enquote{{The small-scale dynamo and the amplification of magnetic
  fields in massive primordial haloes},} \emph{\mnras} \textbf{432},  668--678.

\bibitem[{{Laureijs} \emph{et~al.}(2011){Laureijs}, {Amiaux}, {Arduini},
  {Augu{\`e}res}, {Brinchmann}, {Cole}, {Cropper}, {Dabin}, {Duvet}, {Ealet}
  and et~al.}]{Laureijs2011}
{Laureijs}, R., {Amiaux}, J., {Arduini}, S., {Augu{\`e}res}, J.~.,
  {Brinchmann}, J., {Cole}, R., {Cropper}, M., {Dabin}, C., {Duvet}, L.,
  {Ealet}, A.,  and et~al. (2011). \enquote{{Euclid Definition Study Report},}
  \emph{ArXiv e-prints:1110.3193} .

\bibitem[{{Lazar} \emph{et~al.}(2009){Lazar}, {Schlickeiser}, {Wielebinski} and
  {Poedts}}]{Lazar2009}
{Lazar}, M., {Schlickeiser}, R., {Wielebinski}, R.,  and {Poedts}, S. (2009).
  \enquote{{Cosmological Effects of Weibel-Type Instabilities},} \emph{\apj}
  \textbf{693},  1133--1141.

\bibitem[{{Loeb}(2010)}]{Loeb2010}
{Loeb}, A. (2010). \emph{{How Did the First Stars and Galaxies Form? By Abraham
  Loeb.~Princeton University Press, 2010.~ISBN: 978-1-4008-3406-8}} (Princeton
  University Press).

\bibitem[{{Loeb} and {Furlanetto}(2013)}]{Loeb2013}
{Loeb}, A. and {Furlanetto}, S.~R. (2013). \emph{{The First Galaxies in the
  Universe}} (Princeton University Press).

\bibitem[{{Mac Low} and {Klessen}(2004)}]{Maclow2004}
{Mac Low}, M.-M. and {Klessen}, R.~S. (2004). \enquote{{Control of star
  formation by supersonic turbulence},} \emph{Reviews of Modern Physics}
  \textbf{76},  125--194.

\bibitem[{{Machida} and {Doi}(2013)}]{Machida2013}
{Machida}, M.~N. and {Doi}, K. (2013). \enquote{{The formation of Population
  III stars in gas accretion stage: effects of magnetic fields},} \emph{\mnras}
  \textbf{435},  3283--3305.

\bibitem[{{Machida} \emph{et~al.}(2008){Machida}, {Matsumoto} and
  {Inutsuka}}]{Machida2008}
{Machida}, M.~N., {Matsumoto}, T.,  and {Inutsuka}, S.-i. (2008).
  \enquote{{Magnetohydrodynamics of Population III Star Formation},}
  \emph{\apj} \textbf{685}, 690-704.

\bibitem[{{Machida} \emph{et~al.}(2006){Machida}, {Omukai}, {Matsumoto} and
  {Inutsuka}}]{Machida2006}
{Machida}, M.~N., {Omukai}, K., {Matsumoto}, T.,  and {Inutsuka}, S.-i. (2006).
  \enquote{{The First Jets in the Universe: Protostellar Jets from the First
  Stars},} \emph{\apjl} \textbf{647},  L1--L4.

\bibitem[{{Madau} \emph{et~al.}(2014){Madau}, {Haardt} and {Dotti}}]{Madau2014}
{Madau}, P., {Haardt}, F.,  and {Dotti}, M. (2014). \enquote{{Super-critical
  Growth of Massive Black Holes from Stellar-mass Seeds},} \emph{\apjl}
  \textbf{784}, L38.

\bibitem[{{Maeder} and {Meynet}(2012)}]{Maeder2012}
{Maeder}, A. and {Meynet}, G. (2012). \enquote{{Rotating massive stars: From
  first stars to gamma ray bursts},} \emph{Reviews of Modern Physics}
  \textbf{84},  25--63.

\bibitem[{{Magg} \emph{et~al.}(2018){Magg}, {Hartwig}, {Agarwal}, {Frebel},
  {Glover}, {Griffen} and {Klessen}}]{Magg2017}
{Magg}, M., {Hartwig}, T., {Agarwal}, B., {Frebel}, A., {Glover}, S.~C.~O.,
  {Griffen}, B.~F.,  and {Klessen}, R.~S. (2018). \enquote{{Predicting the
  locations of possible long-lived low-mass first stars: importance of
  satellite dwarf galaxies},} \emph{\mnras} \textbf{473},  5308--5323.

\bibitem[{{Magg} \emph{et~al.}(2016){Magg}, {Hartwig}, {Glover}, {Klessen} and
  {Whalen}}]{Magg2016}
{Magg}, M., {Hartwig}, T., {Glover}, S.~C.~O., {Klessen}, R.~S.,  and {Whalen},
  D.~J. (2016). \enquote{{A new statistical model for Population III supernova
  rates: discriminating between {$\Lambda$}CDM and WDM cosmologies},}
  \emph{\mnras} \textbf{462},  3591--3601.

\bibitem[{{Maio} \emph{et~al.}(2011{\natexlab{a}}){Maio}, {Khochfar}, {Johnson}
  and {Ciardi}}]{Maio2011}
{Maio}, U., {Khochfar}, S., {Johnson}, J.~L.,  and {Ciardi}, B.
  (2011{\natexlab{a}}). \enquote{{The interplay between chemical and mechanical
  feedback from the first generation of stars},} \emph{\mnras} \textbf{414},
  1145--1157.

\bibitem[{{Maio} \emph{et~al.}(2011{\natexlab{b}}){Maio}, {Koopmans} and
  {Ciardi}}]{Maio11}
{Maio}, U., {Koopmans}, L.~V.~E.,  and {Ciardi}, B. (2011{\natexlab{b}}).
  \enquote{{The impact of primordial supersonic flows on early structure
  formation, reionization and the lowest-mass dwarf galaxies},} \emph{\mnras}
  \textbf{412},  L40--L44.

\bibitem[{{Mapelli} \emph{et~al.}(2006){Mapelli}, {Ferrara} and
  {Rea}}]{Mapelli2006}
{Mapelli}, M., {Ferrara}, A.,  and {Rea}, N. (2006). \enquote{{Constraints on
  Galactic intermediate mass black holes},} \emph{\mnras} \textbf{368},
  1340--1350.

\bibitem[{{McDowell}(1961)}]{Mcdowell1961}
{McDowell}, M.~R.~C. (1961). \enquote{{On the formation of H2 in H I regions},}
  \emph{The Observatory} \textbf{81},  240--243.

\bibitem[{{McKee} and {Ostriker}(2007)}]{Mckee2007}
{McKee}, C.~F. and {Ostriker}, E.~C. (2007). \enquote{{Theory of Star
  Formation},} \emph{\araa} \textbf{45},  565--687.

\bibitem[{{McKee} and {Tan}(2008)}]{Mckee2008}
{McKee}, C.~F. and {Tan}, J.~C. (2008). \enquote{{The Formation of the First
  Stars. II. Radiative Feedback Processes and Implications for the Initial Mass
  Function},} \emph{\apj} \textbf{681},  771--797.

\bibitem[{{McQuinn} and {O'Leary}(2012)}]{McQuinn2012}
{McQuinn}, M. and {O'Leary}, R.~M. (2012). \enquote{{The Impact of the
  Supersonic Baryon-Dark Matter Velocity Difference on the z \~{} 20 21 cm
  Background},} \emph{\apj} \textbf{760}, 3.

\bibitem[{{Medvedev} \emph{et~al.}(2004){Medvedev}, {Silva}, {Fiore}, {Fonseca}
  and {Mori}}]{Medvedev2004}
{Medvedev}, M.~V., {Silva}, L.~O., {Fiore}, M., {Fonseca}, R.~A.,  and {Mori},
  W.~B. (2004). \enquote{{Generation of Magnetic Fields in Cosmological
  Shocks},} \emph{Journal of Korean Astronomical Society} \textbf{37},
  533--541.

\bibitem[{{Mortlock} \emph{et~al.}(2011){Mortlock}, {Warren}, {Venemans},
  {Patel}, {Hewett}, {McMahon}, {Simpson}, {Theuns}, {Gonz{\'a}les-Solares} and
  {Adamson, A.~et~al.}}]{Mortlock2011}
{Mortlock}, D.~J., {Warren}, S.~J., {Venemans}, B.~P., {Patel}, M., {Hewett},
  P.~C., {McMahon}, R.~G., {Simpson}, C., {Theuns}, T., {Gonz{\'a}les-Solares},
  E.~A.,  and {Adamson, A.~et~al.} (2011). \enquote{{A luminous quasar at a
  redshift of z = 7.085},} \emph{\nat} \textbf{474},  616--619.

\bibitem[{{Nakama} \emph{et~al.}(2017){Nakama}, {Silk} and
  {Kamionkowski}}]{Nakama2017}
{Nakama}, T., {Silk}, J.,  and {Kamionkowski}, M. (2017). \enquote{{Stochastic
  gravitational waves associated with the formation of primordial black
  holes},} \emph{\prd} \textbf{95}, 4, 043511.

\bibitem[{{Naoz} \emph{et~al.}(2012){Naoz}, {Yoshida} and {Gnedin}}]{Naoz2012}
{Naoz}, S., {Yoshida}, N.,  and {Gnedin}, N.~Y. (2012). \enquote{{Simulations
  of Early Baryonic Structure Formation with Stream Velocity. I. Halo
  Abundance},} \emph{\apj} \textbf{747}, 128.

\bibitem[{{Naoz} \emph{et~al.}(2013){Naoz}, {Yoshida} and {Gnedin}}]{Naoz2013}
{Naoz}, S., {Yoshida}, N.,  and {Gnedin}, N.~Y. (2013). \enquote{{Simulations
  of Early Baryonic Structure Formation with Stream Velocity. II. The Gas
  Fraction},} \emph{\apj} \textbf{763}, 27.

\bibitem[{{Norris} \emph{et~al.}(2013){Norris}, {Yong}, {Bessell},
  {Christlieb}, {Asplund}, {Gilmore}, {Wyse}, {Beers}, {Barklem}, {Frebel} and
  {Ryan}}]{Norris2013}
{Norris}, J.~E., {Yong}, D., {Bessell}, M.~S., {Christlieb}, N., {Asplund}, M.,
  {Gilmore}, G., {Wyse}, R.~F.~G., {Beers}, T.~C., {Barklem}, P.~S., {Frebel},
  A.,  and {Ryan}, S.~G. (2013). \enquote{{The Most Metal-poor Stars. IV. The
  Two Populations with Fe/H < -3.0},} \emph{\apj} \textbf{762}, 28.

\bibitem[{{Offner} \emph{et~al.}(2010){Offner}, {Kratter}, {Matzner},
  {Krumholz} and {Klein}}]{Offner2010}
{Offner}, S.~S.~R., {Kratter}, K.~M., {Matzner}, C.~D., {Krumholz}, M.~R.,  and
  {Klein}, R.~I. (2010). \enquote{{The Formation of Low-mass Binary Star
  Systems Via Turbulent Fragmentation},} \emph{\apj} \textbf{725},  1485--1494.

\bibitem[{{Oguri} and {Marshall}(2010)}]{Oguri2010}
{Oguri}, M. and {Marshall}, P.~J. (2010). \enquote{{Gravitationally lensed
  quasars and supernovae in future wide-field optical imaging surveys},}
  \emph{\mnras} \textbf{405},  2579--2593.

\bibitem[{{O'Leary} and {McQuinn}(2012)}]{Oleary2012}
{O'Leary}, R.~M. and {McQuinn}, M. (2012). \enquote{{The Formation of the First
  Cosmic Structures and the Physics of the z \~{} 20 Universe},} \emph{\apj}
  \textbf{760}, 4.

\bibitem[{{Omukai}(2001)}]{Omukai2001}
{Omukai}, K. (2001). \enquote{{Primordial Star Formation under Far-Ultraviolet
  Radiation},} \emph{\apj} \textbf{546},  635--651.

\bibitem[{{Omukai} and {Nishi}(1998)}]{Omukai1998}
{Omukai}, K. and {Nishi}, R. (1998). \enquote{{Formation of Primordial
  Protostars},} \emph{\apj} \textbf{508},  141--150.

\bibitem[{{Omukai} \emph{et~al.}(2005){Omukai}, {Tsuribe}, {Schneider} and
  {Ferrara}}]{Omukai2005}
{Omukai}, K., {Tsuribe}, T., {Schneider}, R.,  and {Ferrara}, A. (2005).
  \enquote{{Thermal and Fragmentation Properties of Star-forming Clouds in
  Low-Metallicity Environments},} \emph{\apj} \textbf{626},  627--643.

\bibitem[{{O'Shea} and {Norman}(2007)}]{Oshea2007}
{O'Shea}, B.~W. and {Norman}, M.~L. (2007). \enquote{{Population III Star
  Formation in a {$\Lambda$}CDM Universe. I. The Effect of Formation Redshift
  and Environment on Protostellar Accretion Rate},} \emph{\apj} \textbf{654},
  66--92.

\bibitem[{{Padoan} \emph{et~al.}(2014){Padoan}, {Federrath}, {Chabrier},
  {Evans}, {Johnstone}, {J{\o}rgensen}, {McKee} and {Nordlund}}]{Padoan2014}
{Padoan}, P., {Federrath}, C., {Chabrier}, G., {Evans}, N.~J., II, {Johnstone},
  D., {J{\o}rgensen}, J.~K., {McKee}, C.~F.,  and {Nordlund}, {\AA}. (2014).
  \enquote{{The Star Formation Rate of Molecular Clouds},} \emph{Protostars and
  Planets VI} ,  77--100.

\bibitem[{{Palla} \emph{et~al.}(1983){Palla}, {Salpeter} and
  {Stahler}}]{Palla1983}
{Palla}, F., {Salpeter}, E.~E.,  and {Stahler}, S.~W. (1983).
  \enquote{{Primordial star formation - The role of molecular hydrogen},}
  \emph{\apj} \textbf{271},  632--641.

\bibitem[{{Pan} \emph{et~al.}(2012){Pan}, {Loeb} and {Kasen}}]{Pan2012}
{Pan}, T., {Loeb}, A.,  and {Kasen}, D. (2012). \enquote{{Pair-instability
  supernovae via collision runaway in young dense star clusters},}
  \emph{\mnras} \textbf{423},  2203--2208.

\bibitem[{{Peebles} and {Dicke}(1968)}]{Peebles1968}
{Peebles}, P.~J.~E. and {Dicke}, R.~H. (1968). \enquote{{Origin of the Globular
  Star Clusters},} \emph{\apj} \textbf{154},  891.

\bibitem[{{Penston}(1969)}]{Penston1969}
{Penston}, M.~V. (1969). \enquote{{Dynamics of self-gravitating gaseous
  spheres-III. Analytical results in the free-fall of isothermal cases},}
  \emph{\mnras} \textbf{144},  425.

\bibitem[{{Peters} \emph{et~al.}(2011){Peters}, {Banerjee}, {Klessen} and {Mac
  Low}}]{Peters2011}
{Peters}, T., {Banerjee}, R., {Klessen}, R.~S.,  and {Mac Low}, M.-M. (2011).
  \enquote{{The Interplay of Magnetic Fields, Fragmentation, and Ionization
  Feedback in High-mass Star Formation},} \emph{\apj} \textbf{729}, 72.

\bibitem[{{Peters} \emph{et~al.}(2010){Peters}, {Mac Low}, {Banerjee},
  {Klessen} and {Dullemond}}]{Peters2010}
{Peters}, T., {Mac Low}, M.-M., {Banerjee}, R., {Klessen}, R.~S.,  and
  {Dullemond}, C.~P. (2010). \enquote{{Understanding Spatial and Spectral
  Morphologies of Ultracompact H II Regions},} \emph{\apj} \textbf{719},
  831--843.

\bibitem[{{Peters} \emph{et~al.}(2014){Peters}, {Schleicher}, {Smith},
  {Schmidt} and {Klessen}}]{Peters2014}
{Peters}, T., {Schleicher}, D.~R.~G., {Smith}, R.~J., {Schmidt}, W.,  and
  {Klessen}, R.~S. (2014). \enquote{{Low-metallicity star formation: relative
  impact of metals and magnetic fields},} \emph{\mnras} \textbf{442},
  3112--3126.

\bibitem[{{Planck Collaboration} \emph{et~al.}(2016){Planck Collaboration},
  {Ade}, {Aghanim}, {Arnaud}, {Ashdown}, {Aumont}, {Baccigalupi}, {Banday},
  {Barreiro}, {Bartlett} and et~al.}]{Planck2016}
{Planck Collaboration}, {Ade}, P.~A.~R., {Aghanim}, N., {Arnaud}, M.,
  {Ashdown}, M., {Aumont}, J., {Baccigalupi}, C., {Banday}, A.~J., {Barreiro},
  R.~B., {Bartlett}, J.~G.,  and et~al. (2016). \enquote{{Planck 2015 results.
  XIII. Cosmological parameters},} \emph{\aap} \textbf{594}, A13.

\bibitem[{{Pudritz} and {Silk}(1989)}]{Pudritz1989}
{Pudritz}, R.~E. and {Silk}, J. (1989). \enquote{{The origin of magnetic fields
  and primordial stars in protogalaxies},} \emph{\apj} \textbf{342},  650--659.

\bibitem[{{Rafikov}(2001)}]{Rafikov2001}
{Rafikov}, R.~R. (2001). \enquote{{The local axisymmetric instability criterion
  in a thin, rotating, multicomponent disc},} \emph{\mnras} \textbf{323},
  445--452.

\bibitem[{{Ripamonti} and {Abel}(2004)}]{Ripamonti2004}
{Ripamonti}, E. and {Abel}, T. (2004). \enquote{{Fragmentation and the
  formation of primordial protostars: the possible role of collision-induced
  emission},} \emph{\mnras} \textbf{348},  1019--1034.

\bibitem[{{Ripamonti} \emph{et~al.}(2010){Ripamonti}, {Iocco}, {Ferrara},
  {Schneider}, {Bressan} and {Marigo}}]{Ripamonti2010}
{Ripamonti}, E., {Iocco}, F., {Ferrara}, A., {Schneider}, R., {Bressan}, A.,
  and {Marigo}, P. (2010). \enquote{{First star formation with dark matter
  annihilation},} \emph{\mnras} \textbf{406},  2605--2615.

\bibitem[{{Roederer} \emph{et~al.}(2016){Roederer}, {Mateo}, {Bailey}, {Song},
  {Bell}, {Crane}, {Loebman}, {Nidever}, {Olszewski} and {Shectman,
  S.~A.~et~al.}}]{Roederer2016}
{Roederer}, I.~U., {Mateo}, M., {Bailey}, J.~I., III, {Song}, Y., {Bell},
  E.~F., {Crane}, J.~D., {Loebman}, S., {Nidever}, D.~L., {Olszewski}, E.~W.,
  and {Shectman, S.~A.~et~al.} (2016). \enquote{{Detailed Chemical Abundances
  in the r-process-rich Ultra-faint Dwarf Galaxy Reticulum 2},} \emph{\aj}
  \textbf{151}, 82.

\bibitem[{{Rosen} \emph{et~al.}(2016){Rosen}, {Krumholz}, {McKee} and
  {Klein}}]{Rosen2016}
{Rosen}, A.~L., {Krumholz}, M.~R., {McKee}, C.~F.,  and {Klein}, R.~I. (2016).
  \enquote{{An unstable truth: how massive stars get their mass},}
  \emph{\mnras} \textbf{463},  2553--2573.

\bibitem[{{Salvadori} \emph{et~al.}(2010){Salvadori}, {Ferrara}, {Schneider},
  {Scannapieco} and {Kawata}}]{Salvadori2010}
{Salvadori}, S., {Ferrara}, A., {Schneider}, R., {Scannapieco}, E.,  and
  {Kawata}, D. (2010). \enquote{{Mining the Galactic halo for very metal-poor
  stars},} \emph{\mnras} \textbf{401},  L5--L9.

\bibitem[{{Salvadori} \emph{et~al.}(2007){Salvadori}, {Schneider} and
  {Ferrara}}]{Salvadori2007}
{Salvadori}, S., {Schneider}, R.,  and {Ferrara}, A. (2007). \enquote{{Cosmic
  stellar relics in the Galactic halo},} \emph{\mnras} \textbf{381},  647--662.

\bibitem[{{Salvadori} \emph{et~al.}(2015){Salvadori}, {Sk{\'u}lad{\'o}ttir} and
  {Tolstoy}}]{Salvadori2015}
{Salvadori}, S., {Sk{\'u}lad{\'o}ttir}, {\'A}.,  and {Tolstoy}, E. (2015).
  \enquote{{Carbon-enhanced metal-poor stars in dwarf galaxies},} \emph{\mnras}
  \textbf{454},  1320--1331.

\bibitem[{{Santoro} and {Shull}(2006)}]{Santoro2006}
{Santoro}, F. and {Shull}, J.~M. (2006). \enquote{{Critical Metallicity and
  Fine-Structure Emission of Primordial Gas Enriched by the First Stars},}
  \emph{\apj} \textbf{643},  26--37.

\bibitem[{{Sasaki} \emph{et~al.}(2016){Sasaki}, {Suyama}, {Tanaka} and
  {Yokoyama}}]{Sasaki2016}
{Sasaki}, M., {Suyama}, T., {Tanaka}, T.,  and {Yokoyama}, S. (2016).
  \enquote{{Primordial Black Hole Scenario for the Gravitational-Wave Event
  GW150914},} \emph{Physical Review Letters} \textbf{117}, 6, 061101.

\bibitem[{{Saslaw} and {Zipoy}(1967)}]{Saslaw1967}
{Saslaw}, W.~C. and {Zipoy}, D. (1967). \enquote{{Molecular Hydrogen in
  Pre-galactic Gas Clouds},} \emph{\nat} \textbf{216},  976--978.

\bibitem[{{Schaerer}(2002)}]{Schaerer2002}
{Schaerer}, D. (2002). \enquote{{On the properties of massive Population III
  stars and metal-free stellar populations},} \emph{\aap} \textbf{382},
  28--42.

\bibitem[{{Schauer} \emph{et~al.}(2017){Schauer}, {Regan}, {Glover} and
  {Klessen}}]{Schauer17a}
{Schauer}, A.~T.~P., {Regan}, J., {Glover}, S.~C.~O.,  and {Klessen}, R.~S.
  (2017). \enquote{{The formation of direct collapse black holes under the
  influence of streaming velocities},} \emph{\mnras} \textbf{471},  4878--4884.

\bibitem[{{Schekochihin} \emph{et~al.}(2002){Schekochihin}, {Cowley},
  {Hammett}, {Maron} and {McWilliams}}]{Scheko02}
{Schekochihin}, A.~A., {Cowley}, S.~C., {Hammett}, G.~W., {Maron}, J.~L.,  and
  {McWilliams}, J.~C. (2002). \enquote{{A model of nonlinear evolution and
  saturation of the turbulent MHD dynamo},} \emph{New Journal of Physics}
  \textbf{4}, ~84.

\bibitem[{{Schekochihin} \emph{et~al.}(2004){Schekochihin}, {Cowley}, {Taylor},
  {Hammett}, {Maron} and {McWilliams}}]{Schekochihin2004}
{Schekochihin}, A.~A., {Cowley}, S.~C., {Taylor}, S.~F., {Hammett}, G.~W.,
  {Maron}, J.~L.,  and {McWilliams}, J.~C. (2004). \enquote{{Saturated State of
  the Nonlinear Small-Scale Dynamo},} \emph{Physical Review Letters}
  \textbf{92}, 8, 084504.

\bibitem[{{Schleicher} \emph{et~al.}(2010){Schleicher}, {Banerjee}, {Sur},
  {Arshakian}, {Klessen}, {Beck} and {Spaans}}]{Schliecher2010dyn}
{Schleicher}, D.~R.~G., {Banerjee}, R., {Sur}, S., {Arshakian}, T.~G.,
  {Klessen}, R.~S., {Beck}, R.,  and {Spaans}, M. (2010). \enquote{{Small-scale
  dynamo action during the formation of the first stars and galaxies. I. The
  ideal MHD limit},} \emph{\aap} \textbf{522}, A115.

\bibitem[{{Schleicher} \emph{et~al.}(2009){Schleicher}, {Galli}, {Glover},
  {Banerjee}, {Palla}, {Schneider} and {Klessen}}]{Schleicher2009}
{Schleicher}, D.~R.~G., {Galli}, D., {Glover}, S.~C.~O., {Banerjee}, R.,
  {Palla}, F., {Schneider}, R.,  and {Klessen}, R.~S. (2009). \enquote{{The
  Influence of Magnetic Fields on the Thermodynamics of Primordial Star
  Formation},} \emph{\apj} \textbf{703},  1096--1106.

\bibitem[{{Schlickeiser} and {Shukla}(2003)}]{Schlickeiser2003}
{Schlickeiser}, R. and {Shukla}, P.~K. (2003). \enquote{{Cosmological Magnetic
  Field Generation by the Weibel Instability},} \emph{\apjl} \textbf{599},
  L57--L60.

\bibitem[{{Schneider} \emph{et~al.}(2012){Schneider}, {Omukai}, {Limongi},
  {Ferrara}, {Salvaterra}, {Chieffi} and {Bianchi}}]{Schneider2012}
{Schneider}, R., {Omukai}, K., {Limongi}, M., {Ferrara}, A., {Salvaterra}, R.,
  {Chieffi}, A.,  and {Bianchi}, S. (2012). \enquote{{The formation of the
  extremely primitive star SDSS J102915+172927 relies on dust},} \emph{\mnras}
  \textbf{423},  L60--L64.

\bibitem[{{Schneider} \emph{et~al.}(2006){Schneider}, {Salvaterra}, {Ferrara}
  and {Ciardi}}]{Schneider2006}
{Schneider}, R., {Salvaterra}, R., {Ferrara}, A.,  and {Ciardi}, B. (2006).
  \enquote{{Constraints on the initial mass function of the first stars},}
  \emph{\mnras} \textbf{369},  825--834.

\bibitem[{{Schober} \emph{et~al.}(2012{\natexlab{a}}){Schober}, {Schleicher},
  {Federrath}, {Glover}, {Klessen} and {Banerjee}}]{Schobera}
{Schober}, J., {Schleicher}, D., {Federrath}, C., {Glover}, S., {Klessen},
  R.~S.,  and {Banerjee}, R. (2012{\natexlab{a}}). \enquote{{The Small-scale
  Dynamo and Non-ideal Magnetohydrodynamics in Primordial Star Formation},}
  \emph{\apj} \textbf{754}, 99.

\bibitem[{{Schober} \emph{et~al.}(2012{\natexlab{b}}){Schober}, {Schleicher},
  {Federrath}, {Klessen} and {Banerjee}}]{Schoberb}
{Schober}, J., {Schleicher}, D., {Federrath}, C., {Klessen}, R.,  and
  {Banerjee}, R. (2012{\natexlab{b}}). \enquote{{Magnetic field amplification
  by small-scale dynamo action: Dependence on turbulence models and Reynolds
  and Prandtl numbers},} \emph{\pre} \textbf{85}, 2, 026303.

\bibitem[{{Schober} \emph{et~al.}(2015){Schober}, {Schleicher}, {Federrath},
  {Bovino} and {Klessen}}]{Schober2015}
{Schober}, J., {Schleicher}, D.~R.~G., {Federrath}, C., {Bovino}, S.,  and
  {Klessen}, R.~S. (2015). \enquote{{Saturation of the turbulent dynamo},}
  \emph{\pre} \textbf{92}, 2, 023010.

\bibitem[{{Shu}(1977)}]{Shu1977}
{Shu}, F.~H. (1977). \enquote{{Self-similar collapse of isothermal spheres and
  star formation},} \emph{\apj} \textbf{214},  488--497.

\bibitem[{{Shu} \emph{et~al.}(1987){Shu}, {Adams} and {Lizano}}]{Shu1987}
{Shu}, F.~H., {Adams}, F.~C.,  and {Lizano}, S. (1987). \enquote{{Star
  formation in molecular clouds - Observation and theory},} \emph{\araa}
  \textbf{25},  23--81.

\bibitem[{{Sigl} \emph{et~al.}(1997){Sigl}, {Olinto} and {Jedamzik}}]{Sigl1997}
{Sigl}, G., {Olinto}, A.~V.,  and {Jedamzik}, K. (1997). \enquote{{Primordial
  magnetic fields from cosmological first order phase transitions},}
  \emph{\prd} \textbf{55},  4582--4590.

\bibitem[{{Silk}(1986)}]{Silk1986}
{Silk}, J. (1986). \emph{{The cosmic microwave background radiation -
  Implications for galaxy formation}} (University of Chicago Press),  143--158.

\bibitem[{{Silk} and {Langer}(2006)}]{Silk2006}
{Silk}, J. and {Langer}, M. (2006). \enquote{{On the first generation of
  stars},} \emph{\mnras} \textbf{371},  444--450.

\bibitem[{{Skidmore} \emph{et~al.}(2015){Skidmore}, {TMT International Science
  Development Teams} and {Science Advisory Committee}}]{Skidmore2015}
{Skidmore}, W., {TMT International Science Development Teams},  and {Science
  Advisory Committee}, T. (2015). \enquote{{Thirty Meter Telescope Detailed
  Science Case: 2015},} \emph{Research in Astronomy and Astrophysics}
  \textbf{15}, 1945.

\bibitem[{{Smith} \emph{et~al.}(2009){Smith}, {Turk}, {Sigurdsson}, {O'Shea}
  and {Norman}}]{Smith2009}
{Smith}, B.~D., {Turk}, M.~J., {Sigurdsson}, S., {O'Shea}, B.~W.,  and
  {Norman}, M.~L. (2009). \enquote{{Three Modes of Metal-Enriched Star
  Formation in the Early Universe},} \emph{\apj} \textbf{691},  441--451.

\bibitem[{{Smith} \emph{et~al.}(2011){Smith}, {Glover}, {Clark}, {Greif} and
  {Klessen}}]{Smith11}
{Smith}, R.~J., {Glover}, S.~C.~O., {Clark}, P.~C., {Greif}, T.,  and
  {Klessen}, R.~S. (2011). \enquote{{The effects of accretion luminosity upon
  fragmentation in the early universe},} \emph{\mnras} \textbf{414},
  3633--3644.

\bibitem[{{Smith} \emph{et~al.}(2012){Smith}, {Hosokawa}, {Omukai}, {Glover}
  and {Klessen}}]{Smith2012}
{Smith}, R.~J., {Hosokawa}, T., {Omukai}, K., {Glover}, S.~C.~O.,  and
  {Klessen}, R.~S. (2012). \enquote{{Variable accretion rates and fluffy first
  stars},} \emph{\mnras} \textbf{424},  457--463.

\bibitem[{{Sormani} \emph{et~al.}(2017){Sormani}, {Tre{\ss}}, {Klessen} and
  {Glover}}]{Sormani2017}
{Sormani}, M.~C., {Tre{\ss}}, R.~G., {Klessen}, R.~S.,  and {Glover}, S.~C.~O.
  (2017). \enquote{{A simple method to convert sink particles into stars},}
  \emph{\mnras} \textbf{466},  407--412.

\bibitem[{{Spergel} \emph{et~al.}(2015){Spergel}, {Gehrels}, {Baltay},
  {Bennett}, {Breckinridge}, {Donahue}, {Dressler}, {Wallace}, {Whipple},
  {Wollack} and {Zhao}}]{Spergel2015}
{Spergel}, D., {Gehrels}, N., {Baltay}, C., {Bennett}, D., {Breckinridge}, J.,
  {Donahue}, M., {Dressler}, A., {Wallace}, J.~K., {Whipple}, A., {Wollack},
  E.,  and {Zhao}, F. (2015). \enquote{{Wide-Field InfrarRed Survey
  Telescope-Astrophysics Focused Telescope Assets WFIRST-AFTA 2015 Report},}
  \emph{ArXiv e-prints:1503.03757} .

\bibitem[{{Spolyar} \emph{et~al.}(2008){Spolyar}, {Freese} and
  {Gondolo}}]{Spolyar2008}
{Spolyar}, D., {Freese}, K.,  and {Gondolo}, P. (2008). \enquote{{Dark Matter
  and the First Stars: A New Phase of Stellar Evolution},} \emph{Physical
  Review Letters} \textbf{100}, 5, 051101.

\bibitem[{{Stacy} and {Bromm}(2013)}]{Stacy2013}
{Stacy}, A. and {Bromm}, V. (2013). \enquote{{Constraining the statistics of
  Population III binaries},} \emph{\mnras} \textbf{433},  1094--1107.

\bibitem[{{Stacy} \emph{et~al.}(2016){Stacy}, {Bromm} and {Lee}}]{Stacy2016}
{Stacy}, A., {Bromm}, V.,  and {Lee}, A.~T. (2016). \enquote{{Building up the
  Population III initial mass function from cosmological initial conditions},}
  \emph{\mnras} \textbf{462},  1307--1328.

\bibitem[{{Stacy} \emph{et~al.}(2011){Stacy}, {Bromm} and {Loeb}}]{Stacy2011}
{Stacy}, A., {Bromm}, V.,  and {Loeb}, A. (2011). \enquote{{Effect of Streaming
  Motion of Baryons Relative to Dark Matter on the Formation of the First
  Stars},} \emph{\apjl} \textbf{730}, L1.

\bibitem[{{Stacy} \emph{et~al.}(2010){Stacy}, {Greif} and {Bromm}}]{Stacy2010}
{Stacy}, A., {Greif}, T.~H.,  and {Bromm}, V. (2010). \enquote{{The first
  stars: formation of binaries and small multiple systems},} \emph{\mnras}
  \textbf{403},  45--60.

\bibitem[{{Stacy} \emph{et~al.}(2012){Stacy}, {Greif} and {Bromm}}]{Stacy2012}
{Stacy}, A., {Greif}, T.~H.,  and {Bromm}, V. (2012). \enquote{{The first
  stars: mass growth under protostellar feedback},} \emph{\mnras} \textbf{422},
   290--309.

\bibitem[{{Stacy} \emph{et~al.}(2014){Stacy}, {Pawlik}, {Bromm} and
  {Loeb}}]{Stacy2014}
{Stacy}, A., {Pawlik}, A.~H., {Bromm}, V.,  and {Loeb}, A. (2014).
  \enquote{{The mutual interaction between Population III stars and
  self-annihilating dark matter},} \emph{\mnras} \textbf{441},  822--836.

\bibitem[{{Subramanian}(1998)}]{Subramanian1998}
{Subramanian}, K. (1998). \enquote{{Can the turbulent galactic dynamo generate
  large-scale magnetic fields?}} \emph{\mnras} \textbf{294},  718.

\bibitem[{{Sunyaev} and {Zel'dovich}(1970)}]{Sunyaev1970}
{Sunyaev}, R.~A. and {Zel'dovich}, Y.~B. (1970). \enquote{{Small-Scale
  Fluctuations of Relic Radiation},} \emph{\apss} \textbf{7},  3--19.

\bibitem[{{Sur} \emph{et~al.}(2012){Sur}, {Federrath}, {Schleicher}, {Banerjee}
  and {Klessen}}]{Sur2012}
{Sur}, S., {Federrath}, C., {Schleicher}, D.~R.~G., {Banerjee}, R.,  and
  {Klessen}, R.~S. (2012). \enquote{{Magnetic field amplification during
  gravitational collapse - influence of turbulence, rotation and gravitational
  compression},} \emph{\mnras} \textbf{423},  3148--3162.

\bibitem[{{Sur} \emph{et~al.}(2010){Sur}, {Schleicher}, {Banerjee}, {Federrath}
  and {Klessen}}]{Sur2010}
{Sur}, S., {Schleicher}, D.~R.~G., {Banerjee}, R., {Federrath}, C.,  and
  {Klessen}, R.~S. (2010). \enquote{{The Generation of Strong Magnetic Fields
  During the Formation of the First Stars},} \emph{\apjl} \textbf{721},
  L134--L138.

\bibitem[{{Susa}(2013)}]{Susa13}
{Susa}, H. (2013). \enquote{{The Mass of the First Stars},} \emph{\apj}
  \textbf{773}, 185.

\bibitem[{{Susa} \emph{et~al.}(2014){Susa}, {Hasegawa} and {Tominaga}}]{Susa14}
{Susa}, H., {Hasegawa}, K.,  and {Tominaga}, N. (2014). \enquote{{The Mass
  Spectrum of the First Stars},} \emph{\apj} \textbf{792}, 32.

\bibitem[{{Tamai} and {Spyromilio}(2014)}]{Tamai2014}
{Tamai}, R. and {Spyromilio}, J. (2014). \enquote{{European Extremely Large
  Telescope: progress report},} in \emph{Ground-based and Airborne Telescopes
  V}, \emph{\procspie}, Vol. 9145,  91451E, \doi{10.1117/12.2058467}.

\bibitem[{{Tan} and {McKee}(2004)}]{Tan2004}
{Tan}, J.~C. and {McKee}, C.~F. (2004). \enquote{{The Formation of the First
  Stars. I. Mass Infall Rates, Accretion Disk Structure, and Protostellar
  Evolution},} \emph{\apj} \textbf{603},  383--400.

\bibitem[{{Toomre}(1964)}]{Toomre1964}
{Toomre}, A. (1964). \enquote{{On the gravitational stability of a disk of
  stars},} \emph{\apj} \textbf{139},  1217--1238.

\bibitem[{{Tseliakhovich} \emph{et~al.}(2011){Tseliakhovich}, {Barkana} and
  {Hirata}}]{Tseliakhovich2011}
{Tseliakhovich}, D., {Barkana}, R.,  and {Hirata}, C.~M. (2011).
  \enquote{{Suppression and spatial variation of early galaxies and
  minihaloes},} \emph{\mnras} \textbf{418},  906--915.

\bibitem[{{Tseliakhovich} and {Hirata}(2010)}]{Tseliakhovich2010}
{Tseliakhovich}, D. and {Hirata}, C. (2010). \enquote{{Relative velocity of
  dark matter and baryonic fluids and the formation of the first structures},}
  \emph{\prd} \textbf{82}, 8, 083520.

\bibitem[{{Tsuribe} and {Omukai}(2006)}]{Tsuribe2006}
{Tsuribe}, T. and {Omukai}, K. (2006). \enquote{{Dust-cooling-induced
  Fragmentation of Low-Metallicity Clouds},} \emph{\apjl} \textbf{642},
  L61--L64.

\bibitem[{{Tsuribe} and {Omukai}(2008)}]{Tsuribe2008}
{Tsuribe}, T. and {Omukai}, K. (2008). \enquote{{Physical Mechanism for the
  Intermediate Characteristic Stellar Mass in Extremely Metal Poor
  Environments},} \emph{\apjl} \textbf{676},  L45--L48.

\bibitem[{{Tumlinson}(2006)}]{Tumlinson2006}
{Tumlinson}, J. (2006). \enquote{{Chemical Evolution in Hierarchical Models of
  Cosmic Structure. I. Constraints on the Early Stellar Initial Mass
  Function},} \emph{\apj} \textbf{641},  1--20.

\bibitem[{{Tumlinson}(2010)}]{Tumlinson2010}
{Tumlinson}, J. (2010). \enquote{{Chemical Evolution in Hierarchical Models of
  Cosmic Structure. II. The Formation of the Milky Way Stellar Halo and the
  Distribution of the Oldest Stars},} \emph{\apj} \textbf{708},  1398--1418.

\bibitem[{{Turk} \emph{et~al.}(2009){Turk}, {Abel} and {O'Shea}}]{Turk09}
{Turk}, M.~J., {Abel}, T.,  and {O'Shea}, B. (2009). \enquote{{The Formation of
  Population III Binaries from Cosmological Initial Conditions},}
  \emph{Science} \textbf{325},  601--.

\bibitem[{{Turk} \emph{et~al.}(2012){Turk}, {Oishi}, {Abel} and
  {Bryan}}]{Turk2012}
{Turk}, M.~J., {Oishi}, J.~S., {Abel}, T.,  and {Bryan}, G.~L. (2012).
  \enquote{{Magnetic Fields in Population III Star Formation},} \emph{\apj}
  \textbf{745}, 154.

\bibitem[{{Turk} \emph{et~al.}(2011){Turk}, {Smith}, {Oishi}, {Skory},
  {Skillman}, {Abel} and {Norman}}]{Turk2011}
{Turk}, M.~J., {Smith}, B.~D., {Oishi}, J.~S., {Skory}, S., {Skillman}, S.~W.,
  {Abel}, T.,  and {Norman}, M.~L. (2011). \enquote{{yt: A Multi-code Analysis
  Toolkit for Astrophysical Simulation Data},} \emph{\apjs} \textbf{192}, 9.

\bibitem[{{Umeda} \emph{et~al.}(2016){Umeda}, {Hosokawa}, {Omukai} and
  {Yoshida}}]{Umeda2016}
{Umeda}, H., {Hosokawa}, T., {Omukai}, K.,  and {Yoshida}, N. (2016).
  \enquote{{The Final Fates of Accreting Supermassive Stars},} \emph{\apjl}
  \textbf{830}, L34.

\bibitem[{{Visbal} \emph{et~al.}(2012){Visbal}, {Barkana}, {Fialkov},
  {Tseliakhovich} and {Hirata}}]{Visbal2012}
{Visbal}, E., {Barkana}, R., {Fialkov}, A., {Tseliakhovich}, D.,  and {Hirata},
  C.~M. (2012). \enquote{{The signature of the first stars in atomic hydrogen
  at redshift 20},} \emph{\nat} \textbf{487},  70--73.

\bibitem[{{von Weizs{\"a}cker}(1951)}]{VonWeiz1951}
{von Weizs{\"a}cker}, C.~F. (1951). \enquote{{The Evolution of Galaxies and
  Stars.}} \emph{\apj} \textbf{114},  165.

\bibitem[{{Whalen} \emph{et~al.}(2004){Whalen}, {Abel} and
  {Norman}}]{Whalen2004}
{Whalen}, D., {Abel}, T.,  and {Norman}, M.~L. (2004). \enquote{{Radiation
  Hydrodynamic Evolution of Primordial H II Regions},} \emph{\apj}
  \textbf{610},  14--22.

\bibitem[{{Whalen} \emph{et~al.}(2013{\natexlab{a}}){Whalen}, {Even}, {Frey},
  {Smidt}, {Johnson}, {Lovekin}, {Fryer}, {Stiavelli}, {Holz}, {Heger},
  {Woosley} and {Hungerford}}]{Whalen2013b}
{Whalen}, D.~J., {Even}, W., {Frey}, L.~H., {Smidt}, J., {Johnson}, J.~L.,
  {Lovekin}, C.~C., {Fryer}, C.~L., {Stiavelli}, M., {Holz}, D.~E., {Heger},
  A., {Woosley}, S.~E.,  and {Hungerford}, A.~L. (2013{\natexlab{a}}).
  \enquote{{Finding the First Cosmic Explosions. I. Pair-instability
  Supernovae},} \emph{\apj} \textbf{777}, 110.

\bibitem[{{Whalen} \emph{et~al.}(2013{\natexlab{b}}){Whalen}, {Johnson},
  {Smidt}, {Meiksin}, {Heger}, {Even} and {Fryer}}]{Whalen2013}
{Whalen}, D.~J., {Johnson}, J.~L., {Smidt}, J., {Meiksin}, A., {Heger}, A.,
  {Even}, W.,  and {Fryer}, C.~L. (2013{\natexlab{b}}). \enquote{{The Supernova
  that Destroyed a Protogalaxy: Prompt Chemical Enrichment and Supermassive
  Black Hole Growth},} \emph{\apj} \textbf{774}, 64.

\bibitem[{{Whitworth} and {Summers}(1985)}]{Whitworth1985}
{Whitworth}, A. and {Summers}, D. (1985). \enquote{{Self-similar condensation
  of spherically symmetric self-gravitating isothermal gas clouds},}
  \emph{\mnras} \textbf{214},  1--25.

\bibitem[{{Widrow} \emph{et~al.}(2012){Widrow}, {Ryu}, {Schleicher},
  {Subramanian}, {Tsagas} and {Treumann}}]{Widrow2012}
{Widrow}, L.~M., {Ryu}, D., {Schleicher}, D.~R.~G., {Subramanian}, K.,
  {Tsagas}, C.~G.,  and {Treumann}, R.~A. (2012). \enquote{{The First Magnetic
  Fields},} \emph{\ssr} \textbf{166},  37--70.

\bibitem[{{Wise} and {Abel}(2007)}]{Wise2007}
{Wise}, J.~H. and {Abel}, T. (2007). \enquote{{Resolving the Formation of
  Protogalaxies. I. Virialization},} \emph{\apj} \textbf{665},  899--910.

\bibitem[{{Wise} \emph{et~al.}(2012{\natexlab{a}}){Wise}, {Abel}, {Turk},
  {Norman} and {Smith}}]{Wise2012b}
{Wise}, J.~H., {Abel}, T., {Turk}, M.~J., {Norman}, M.~L.,  and {Smith}, B.~D.
  (2012{\natexlab{a}}). \enquote{{The birth of a galaxy - II. The role of
  radiation pressure},} \emph{\mnras} \textbf{427},  311--326.

\bibitem[{{Wise} \emph{et~al.}(2008){Wise}, {Turk} and {Abel}}]{Wise2008}
{Wise}, J.~H., {Turk}, M.~J.,  and {Abel}, T. (2008). \enquote{{Resolving the
  Formation of Protogalaxies. II. Central Gravitational Collapse},} \emph{\apj}
  \textbf{682},  745--757.

\bibitem[{{Wise} \emph{et~al.}(2012{\natexlab{b}}){Wise}, {Turk}, {Norman} and
  {Abel}}]{Wise2012}
{Wise}, J.~H., {Turk}, M.~J., {Norman}, M.~L.,  and {Abel}, T.
  (2012{\natexlab{b}}). \enquote{{The Birth of a Galaxy: Primordial Metal
  Enrichment and Stellar Populations},} \emph{\apj} \textbf{745}, 50.

\bibitem[{{Woods} \emph{et~al.}(2017){Woods}, {Heger}, {Whalen},
  {Haemmerl{\'e}} and {Klessen}}]{Woods2017}
{Woods}, T.~E., {Heger}, A., {Whalen}, D.~J., {Haemmerl{\'e}}, L.,  and
  {Klessen}, R.~S. (2017). \enquote{{On the Maximum Mass of Accreting
  Primordial Supermassive Stars},} \emph{\apjl} \textbf{842}, L6.

\bibitem[{{Wu} \emph{et~al.}(2015){Wu}, {Wang}, {Fan}, {Yi}, {Zuo}, {Bian},
  {Jiang}, {McGreer}, {Wang}, {Yang}, {Yang}, {Thompson} and {Beletsky}}]{Wu15}
{Wu}, X.-B., {Wang}, F., {Fan}, X., {Yi}, W., {Zuo}, W., {Bian}, F., {Jiang},
  L., {McGreer}, I.~D., {Wang}, R., {Yang}, J., {Yang}, Q., {Thompson}, D.,
  and {Beletsky}, Y. (2015). \enquote{{An ultraluminous quasar with a
  twelve-billion-solar-mass black hole at redshift 6.30},} \emph{Nature}
  \textbf{518},  512--515.

\bibitem[{{Yoon} \emph{et~al.}(2008){Yoon}, {Iocco} and {Akiyama}}]{Yoon2008}
{Yoon}, S.-C., {Iocco}, F.,  and {Akiyama}, S. (2008). \enquote{{Evolution of
  the First Stars with Dark Matter Burning},} \emph{\apjl} \textbf{688}, L1.

\bibitem[{{Yorke} and {Sonnhalter}(2002)}]{Yorke2002}
{Yorke}, H.~W. and {Sonnhalter}, C. (2002). \enquote{{On the Formation of
  Massive Stars},} \emph{\apj} \textbf{569},  846--862.

\bibitem[{{Yoshida} \emph{et~al.}(2003){Yoshida}, {Abel}, {Hernquist} and
  {Sugiyama}}]{Yoshida2003}
{Yoshida}, N., {Abel}, T., {Hernquist}, L.,  and {Sugiyama}, N. (2003).
  \enquote{{Simulations of Early Structure Formation: Primordial Gas Clouds},}
  \emph{\apj} \textbf{592},  645--663.

\bibitem[{{Yoshida} \emph{et~al.}(2012){Yoshida}, {Hosokawa} and
  {Omukai}}]{Yoshida2012}
{Yoshida}, N., {Hosokawa}, T.,  and {Omukai}, K. (2012). \enquote{{Formation of
  the first stars in the universe},} \emph{Progress of Theoretical and
  Experimental Physics} \textbf{2012}, 1, 01A305.

\bibitem[{{Yoshida} \emph{et~al.}(2007){Yoshida}, {Oh}, {Kitayama} and
  {Hernquist}}]{Yoshida2007}
{Yoshida}, N., {Oh}, S.~P., {Kitayama}, T.,  and {Hernquist}, L. (2007).
  \enquote{{Early Cosmological H II/He III Regions and Their Impact on
  Second-Generation Star Formation},} \emph{\apj} \textbf{663},  687--707.

\bibitem[{{Yoshida} \emph{et~al.}(2008){Yoshida}, {Omukai} and
  {Hernquist}}]{Yoshida2008}
{Yoshida}, N., {Omukai}, K.,  and {Hernquist}, L. (2008). \enquote{{Protostar
  Formation in the Early Universe},} \emph{Science} \textbf{321},  669.

\bibitem[{{Yoshida} \emph{et~al.}(2006){Yoshida}, {Omukai}, {Hernquist} and
  {Abel}}]{Yoshida2006}
{Yoshida}, N., {Omukai}, K., {Hernquist}, L.,  and {Abel}, T. (2006).
  \enquote{{Formation of Primordial Stars in a {$\Lambda$}CDM Universe},}
  \emph{\apj} \textbf{652},  6--25.

\bibitem[{{Zinnecker} and {Yorke}(2007)}]{Zinnecker2007}
{Zinnecker}, H. and {Yorke}, H.~W. (2007). \enquote{{Toward Understanding
  Massive Star Formation},} \emph{\araa} \textbf{45},  481--563.

\end{thebibliography}
